\documentclass[%
 reprint,
 amsmath,amssymb,
 aps,
]{revtex4-2}

\usepackage{graphicx}
\usepackage{dcolumn}	
\usepackage{hyperref}
\usepackage{bm}
\usepackage{wasysym}
\usepackage[utf8]{inputenc}
\usepackage[T1]{fontenc}
\usepackage{etoolbox}
\usepackage{titlesec}
\usepackage{subcaption}
\usepackage[dvipsnames]{xcolor}
\usepackage{amsmath}
\usepackage{dsfont}
\usepackage{indentfirst}
\usepackage{xcolor}
\usepackage{amssymb}
\usepackage{mathtools}
\usepackage{bbold}
\usepackage{accents}
\usepackage{multirow}
\usepackage{palatino} 
\usepackage{lmodern}
\usepackage{soul}
\usepackage{svg}
\usepackage{aas_macros}
\usepackage[margin=0.5in]{geometry}
\usepackage{enumitem}
\usepackage{lineno}

\begin{document}

\title{Validating a main beam treatment of parametric, pixel-based component separation in the context of CMB observations}
\author{Arianna Rizzieri$^{1,2}$}
\email{rizzieri@apc.in2p3.fr}
\author{Josquin Errard$^{1}$}
\author{Radek Stompor$^{1,2}$}
\affiliation{$^{1}$ Université Paris Cité, CNRS, Astroparticule et Cosmologie, F-75013 Paris, France}
\affiliation{$^{2}$ CNRS-UCB International Research Laboratory, Centre Pierre Bin\'{e}truy, IRL2007, CPB-IN2P3, Berkeley, US}
\date{\today}

\begin{abstract}
    We implement a simple, main beam correction in the maximum-likelihood, parametric component separation approach, which allows on accounting for different beamwidths of input maps at different frequencies without any preprocessing. We validate the approach on full-sky and cut-sky simulations and discuss the importance and impact of the assumptions and simplifications. 
    We find that, in the cases when the underlying sky model is indeed parametric, the method successfully recovers component spectral parameters and component maps at the pre-defined resolution.
    The improvement on the precision of the estimated spectral parameters is found to be minor due to the redness of the foreground angular spectra, however the method is potentially more accurate, in particular if the foreground properties display strong, spatial variability, as it does not assume commutation of the beam smoothing and mixing matrix operators. The method permits a reconstruction of the CMB map with a resolution significantly superior to that of the lowest resolution map used in the analysis and with the nearly optimal noise level, facilitating exploitation of the cosmological information contained on angular scales, which would be otherwise inaccessible.
    The method preserves all the advantages of a pixel-domain implementation of the parametric approach, and, as it deals with the beams in the harmonic domain, it can also straightforwardly account for spatially stationary map-domain noise correlations.
\end{abstract}

\maketitle


\section{Introduction}\label{sec:intro}
Detection and characterization of the primordial $B$-mode signatures in the polarization of the cosmic microwave background (CMB) anisotropies is one of the major goals of current research in cosmology as
it offers potentially groundbreaking insights into cosmology and fundamental physics. 
A major effort is therefore on-going in the field aimed at achieving this goal.
In this context the large angular scales are both particularly interesting and challenging. 
The interest is due to the fact that some of the most popular models of the origins of the observable Universe tend to predict large angular scale $B$-mode polarization due to primordial gravitational waves they generically produce. 
The challenges are related to the smallness of the expected signals of interest, as compared to signals of other origins.  
Among those, astrophysical contaminations affecting polarization observations throughout the microwave band, stand out as one of the major obstacles. 
These are generically called foregrounds and are mainly due to galactic emissions either by dust particles perpendicularly aligned with the magnetic field of our Galaxy or by free electrons spiraling in that same field, referred to hereafter as dust and synchrotron signals, respectively. 
Removal of these contributions constitutes one of the major steps down the data analysis pipeline of any modern CMB experiment. 
The corresponding procedures are referred to as component separation. 
A number of different component separation approaches have been proposed and validated on simulated and actual data, e.g.~\cite{eriksen2004foreground, cardoso2008component, baccigalupi2000neural, katayama2011simple, brandt1994separation, eriksen2006cosmic, eriksen2008joint, stompor2009maximum, stompor2016forecasting, de2020detection}. 
The different methods make different assumptions concerning the properties of the underlying signals, and include approaches which make no explicit assumptions about the frequency dependence of the foreground components, generically called non-parametric methods, e.g.~\cite{remazeilles2011foreground, remazeilles2021peeling, carones2023multiclustering, leloup2023nonparametric, morshed2024pixel}, as well as methods where the frequency scaling of all components is assumed to be known down to a few parameters, as in the parametric approaches, e.g.~\cite{ichiki2019delta, azzoni2021minimal, vacher2022moment, de2022testing, galloway2023beyondplanck}. 
A successful approach has to be also sufficiently flexible to include instrumental effects, which while originated from the instrument may affect the outcome of the component separation procedures potentially biasing the obtained results~\cite{verges2021framework, jost2023characterizing, andersen2023beyondplanck}.
In this work we focus on accounting on the frequency dependent experimental beams in the context of the parametric component separation technique as introduced in the next section. 
Specifically, we propose a simple approach accounting for main beam differences between different single frequency maps, constituting the input data for the component separation step, directly in the component separation procedure. 
We demonstrate the approach on nearly full-sky and cut sky cases with characteristics motivated by forthcoming $LiteBIRD$~\cite{litebird2023probing} and Simons Observatory ($SO$) surveys~\cite{ade2019simons}, respectively. 
We show that the proposed technique can result in a significantly better resolution of the derived maps of the components, while achieving nearly optimal noise levels and being more robust with respect to systematic biases due to spatial variability of the foreground properties. 
These improvements come however at increased computational cost, we introduce and discuss novel, specialized numerical algorithms which make the implementation of the method feasible.

In the remaining part of Section~\ref{sec:intro} we introduce the component separation formalism, Section~\ref{sec:formalism}, and we discuss the possible data model extensions to account for the beams, Section~\ref{sec:prob_of_beams}.
In Section~\ref{sec:method}, we present the component separation formalism extended to account for the beams and its implementation. 
We then present, in Section~\ref{sec:results}, some results from applying the proposed methods on simulated data emphasizing their differences.
In Section~\ref{sec:conclusion}, we make some concluding remarks and discuss the future prospects.

\subsection{Parametric component separation formalism}\label{sec:formalism}
We assume hereafter that the data collected by the experiment can be cast as a set of sky maps, reflecting the signal as seen at different frequencies and with well-characterized uncertainties. 
This is indeed the case of the scanning experiments, a category which includes the majority of on-going and forthcoming experimental efforts. 
We will also assume that the signal measured at each frequency can be represented as a linear mixture of sky templates each corresponding to a signal due to a different physical source, and referred to as component, as observed at some predefined fiducial frequency. 
This results in the following model of the input data,
\begin{equation}\label{data_model_no_beams}
    \mathbf{d} = \mathbf{A} \mathbf{s} + \mathbf{n}  
\end{equation}
where $\mathbf{d}$ is the data vector combining all the maps at each frequency band and for each Stokes parameter, $\mathbf{A}$ is the component mixing matrix, and $\mathbf{s}$ stands for the templates of the actual signals for each component at the fiducial frequency. $\mathbf{n}$ is the statistical uncertainty assigned to the data assumed to be Gaussian-distributed and due to the instrumental noise.  
The objective of the component separation procedure is to provide reliable estimates of all (or at least some) sky components, $\mathbf{s}$. 
The challenge is due to the fact that neither the mixing matrix, $\mathbf{A}$, nor the component templates are known a priori and have to be both derived from the noisy data.

In the parametric methods, the frequency dependence of the component amplitudes, as encoded by the columns of the mixing matrix, $\mathbf{A}$, is assumed to be known down to a few unknown parameters, $\beta_i$.  For the parametric methods we therefore have $\mathbf{A} = \mathbf{A}(\beta_i)$. These parameters are called spectral parameters and are derived from the data as part of the parametric component separation procedure.
Many existing parametric component separation techniques are based on the maximum likelihood principle, e.g. \cite{eriksen2006cosmic, stompor2009maximum, de2020detection}.

The quantities appearing in Eq.~(\ref{data_model_no_beams}) can be defined in different domains, such as pixel, harmonic, or wavelet.
The choice of domain will in general depend on the specific context, defined by the expected properties of the foregrounds, as well as the specificity of the chosen component separation method. The parametric approach can be implemented in any of these domains.
Among these, the pixel domain offers certain advantages, as this is the domain in which accounting on the inhomogeneous noise and spatial dependence of the foreground properties is the most straightforward. 
The former feature is common to essentially all experiments and is related to the non-uniform scanning of the sky, and the latter is well confirmed observationally and particularly relevant for experiments targeting large patches of the sky as for example is the case of satellite missions. This choice has however a number of downsides, one of which is accounting for instrumental beams. We discuss this problem in detail below.

The consideration presented in this paper are fully general and applicable to any parametric component separation method. However, for concreteness, we follow the implementation of~\cite{stompor2009maximum} and assume that the procedure is implemented in two steps. The first, non-linear step renders the estimates of the spectral parameters, via a maximization of the so-called spectral likelihood,
\begin{eqnarray}
-2 \mathrm{ln} \mathcal{L}_{sp} (\beta_i) 
     & = &  \mathrm{const}
      - \mathbf{d}^T \mathbf{N}^{-1} \mathbf{A} (\mathbf{A}^T \mathbf{N}^{-1} \mathbf{A})^{-1}\mathbf{A}^T \mathbf{N}^{-1} \mathbf{d},\ \ \ 
\label{eq:spec-like-def}
\end{eqnarray}
where $\mathbf{N}$ is the noise covariance matrix.
The second, linear step estimates the sky components from the input data given the precomputed parameters as,
\begin{eqnarray}
  \mathbf{s} & = & (\mathbf{A}^T \mathbf{N}^{-1} \mathbf{A})^{-1} \mathbf{A}^T \mathbf{N}^{-1} \mathbf{d}.
\label{eq:map-make-step}
\end{eqnarray}

\subsection{Accounting for beams}\label{sec:prob_of_beams}
The instrumental beams generically characterize the response of the instrument to sky signals.
A variety of issues related to instrumental beams are relevant in the data analysis of CMB experiments, first of which the modeling and characterization of the beams themselves.
However, our goal here is to deal with beams at the map-level in the component separation step, where multiple frequency maps each with its specific, and potentially different beam, need to be processed simultaneously.
We need to account for the different beams to avoid biases in the component separation procedure, and to determine the beam width of the recovered component maps.

\subsubsection{Basic data model}
The instantaneous beams can be generally very complex, displaying both complicated spatial and frequency dependencies and resulting in complex, pixel-dependent effective beams in the recovered sky maps.
While many of these effects may need to be accounted for in a complete analysis, our focus here is on the main property of the beams, namely, the dependence of their main lobe on frequency bands.
We thus assume, hereafter, that such sufficiently accurate effective beams exist and that they are independent on the sky pixel location.

Under these assumptions the input data set can be modeled as,
    \begin{equation}\label{data_model_beams}
        \mathbf{d} = \mathbf{B_{true}}(\mathbf{A} \mathbf{s}) + \mathbf{n}  
    \end{equation}
where the operator $\mathbf{B}_\mathbf{true}$ represents the convolution with the true beams. 
This operator is a sparse matrix, which is block-diagonal performing the convolutions for each frequency map separately. 
We note that while the matrices $\mathbf{B}_\mathbf{true}$ and $\mathbf{A}$ appear in the data model in a similar manner, they act differently on the maps, with the mixing matrix, $\mathbf{A}$, combining amplitudes of different components corresponding to the same pixel, and the beam matrix, $\mathbf{B}_\mathbf{true}$, combining pixels of the same map.
In the case of a full sky coverage, the application of the beam operator to a sky map would correspond to a multiplication of their representations in the harmonic space. This observation is the basis of the (approximate) algorithm proposed below. 

\subsubsection{Commutation of component mixing and beam smoothing}\label{sec:commutation}

\label{commutation}
Let us assume that the beams for all considered frequency maps are the same and name the corresponding beam operator $\mathbf{B}$. In this case, we can wonder whether convolving the single frequency maps with the corresponding beam operator, $\mathbf{B}$, is equivalent to convolving first the component maps with the same beam and combining them together by applying the mixing matrix. Whenever this is the case, on defining $\mathbf{\hat B}$ as the beam operator acting on the component maps, we can write,
\begin{equation}
    \mathbf{B} \, \mathbf{A} \; = \; \mathbf{A} \, \mathbf{\hat{B}},
     \label{eqn:BAcommutation}
\end{equation}
as the beam smoothing and the component mixing operations can be swapped. We refer to such a case as a beam smoothing and mixing matrix commutation.

While for the commutation to be valid, the experimental beams have to be the same for all the channels, this is not always sufficient, it can be strictly realized only in two situations. First, when the mixing matrix does not have spatial variability, thus it is the same for all the pixels,
and, second, when the component scaling laws depend on both frequency and pixel but the dependence can be factorized into products of two functions depending either on frequency or on pixel number. In this latter case, the corresponding mixing matrix can be represented as a product of a pixel independent matrix and a diagonal matrix defining the scaling-law pixel dependence for each of the foreground components.

However, even if $\mathbf{A}$ is spatially varying in a more general form, we can still hope that a generalized commutation holds which reads, 
    \begin{equation}
        \mathbf{B}\,\mathbf{A} \; = \; \mathbf{A'}\, \mathbf{\hat{B}},   
        \label{eqn:BAcommutation_generlized}
    \end{equation}
where we have defined modified (effective) mixing matrix $\mathbf{A'}$ accounting for the effect of the beam convolution. We note that while there are no generic prescriptions on how $\mathbf{A}$ should be generalized to $\mathbf{A'}$,
a possible approach has been suggested in~\cite{chluba2017rethinking, vacher2023high}.
In the following we do not consider this option, instead we discuss the implications of assuming the commutation as in Eq.~(\ref{eqn:BAcommutation}) and evaluate the error this entails in case of a spatially varying mixing matrix.
We note that the formalism presented below could be straightforwardly generalized to include the cases as in Eq.~(\ref{eqn:BAcommutation_generlized}), if needed.

Practical examples where this commutation is relevant are discussed in the next section.

\subsubsection{Specialized data models}

There are inherently two different manners in which we can proceed starting from Eq.~(\ref{data_model_beams}).
The first one is to smooth all the single frequency maps to some common resolution prior to the component separation, as done in e.g.~\cite{litebird2023probing, wolz2023simons}.
In the following we refer to this method as \textit{common resolution approach (CRA)} and use it as a reference method to compare other methods against.
On denoting the corresponding beam operator as $\mathbf{B}_\mathbf{final}$, we transform the data set modeled as in Eq.~(\ref{data_model_beams}), as follows,
 \begin{eqnarray}
        \mathbf{d'} & \equiv & \mathbf{B}^{\phantom{-1}}_\mathbf{final}\,\mathbf{B}_\mathbf{true}^{-1}\,\mathbf{d} \nonumber \\
        & = & \mathbf{B}_\mathbf{final} \,\mathbf{A} \,\mathbf{s}  + \mathbf{B}_\mathbf{final}\,\mathbf{B}^{-1}_\mathbf{true}\, \mathbf{n}\nonumber \\
& = & \mathbf{B}_\mathbf{final} \,\mathbf{A} \,\mathbf{\hat B}_\mathbf{final}^{-1}\,\mathbf{s}_\mathbf{final}  + \mathbf{B}_\mathbf{final}\,\mathbf{B}^{-1}_\mathbf{true}\, \mathbf{n}\nonumber\\
 & = & \mathbf{A} \mathbf{s}^{\phantom{-1}}_\mathbf{final}  + \mathbf{B}^{\phantom{-1}}_\mathbf{final}\,\mathbf{B}^{-1}_\mathbf{true}\,\mathbf{n}.
        \label{preprocessing_data}
    \end{eqnarray}
Here, $\mathbf{s}_\mathbf{final}$ stands for the component templates all smoothed with the same final beams. We also assume that the inverse beam operator, $\mathbf{\hat B}_\mathbf{final}^{-1}$, is well-defined. This may not always be straightforward given that all the sky maps available in practice are pixelized and some regularization of this operator may be needed in actual applications. We found this to be in practice easily achievable.

The last equality in Eq.~(\ref{preprocessing_data}) assumes that $\mathbf{B}_\mathbf{final} \,\mathbf{A} \,\mathbf{\hat B}_\mathbf{final}^{-1} = \mathbf{A}$ and therefore a commutation of the final beam matrix and the mixing matrix as in Eq.~(\ref{eqn:BAcommutation}).
As discussed in the previous section, this is strictly valid only in case of pixel independent mixing matrix, or for a mixing matrix which can be written as the product of a pixel-independent mixing matrix and a diagonal matrix giving the frequency-independent pixel dependence for each foreground component.
We note that the latter case is mathematically equivalent to the case of the pixel-independent mixing matrix, though the foreground components recovered at the end of the component separation procedure would be products of the actual components and the pixel-dependent functions describing the scaling law dependencies. This will have, however, no impact on the CMB signal recovery procedure. For this reason, for definiteness, we assume hereafter that the mixing matrices do not allow for such a representation. This is indeed the case for the standard foreground scaling laws when their pixel dependence is introduced via a dependence of the spectral parameters on the sky pixel.

In all other cases, the effective mixing matrix given by $\mathbf{B}_\mathbf{final} \,\mathbf{A} \,\mathbf{\hat B}_\mathbf{final}^{-1}$ will not only be pixel-dependent but the resulting frequency scaling of each component will be modified in a way depending on the initial scaling laws but also their pixel-dependence, i.e., as in Eq.~(\ref{eqn:BAcommutation_generlized}).

We note that in the case of the mixing matrices which are effectively constant over the range of angular scales much larger than the involved beam scales, the commutation can be fulfilled approximately and with accuracy progressively improving for smaller beams. In contrast, failure to satisfy the commutation will indicate that the data model assumed in Eq.~(\ref{preprocessing_data}) is insufficient and may lead to an uncontrollable level of biases in the final result.

The components in the new data set, $\mathbf{d'}$, Eq.~(\ref{preprocessing_data}), are now combined together as defined by the mixing matrix without involving any beam operators, as assumed in the standard component separation techniques. 
The beam operations are now either hidden in the component templates, which are now smoothed with the effective beam, as well as in the properties of the map-level noise, which is now both enhanced and correlated. 
Both these effects would have to be taken into account on the first step of the component separation, as any mischaracterization of the noise may lead to significant biases in the estimated spectral parameters, Eq.~(\ref{eq:spec-like-def}).
In the common beam approaches these are typically 
mitigated by first smoothing all the input maps to the common resolution as given by the largest experimental beam, i.e., usually that of the lowest frequency band, and subsequently downgrading the map resolution via increasing the pixel size. The pixel-pixel noise correlation induced by the beam smoothing are consequently suppressed and the effective noise can be considered (nearly) uncorrelated, if the final pixel size is sufficiently large.
The downgraded data can be then described with the data model, 
\begin{equation}
    \mathbf{d'} = \mathbf{A} \mathbf{s}^{\phantom{-1}}_\mathbf{final} +\mathbf{n}^{\phantom{-1}}_\mathbf{eff},
    \label{data_model_common_step1}
\end{equation}
where $\mathbf{n}_\mathbf{eff}$ differs from $\mathbf{n}$ as its level reflects the downgrading procedure and which, by construction, can be presumed to be uncorrelated.
We emphasize that the smoothing/downgrading procedure has to be applied to all input, single frequency maps consistently. We also assume implicitly that it commutes with the mixing matrix. Indeed, given that the downgrade has to be performed on scales larger than the final beam, this may set requirements on spatial variability of the mixing more demanding than the beam commutation itself as already assumed earlier. We also note that while it is possible to use these maps to perform the second step of the component separation procedure, Eq.~(\ref{eq:map-make-step}), this may not be desirable as the resulting component maps, will have complex smoothing determined by the pixelization. Indeed, typically the mixing matrix estimates derived from the smoothed and underpixelized maps are instead used in Eq.~(\ref{eq:map-make-step}) together with the smoothed but not underpixelized input maps. The resolution of the resulting component maps is then determined by the adopted common beam. This procedure assumes usually simple diagonal noises. This is justifiable given the linearity of this step ensuring that the penalty for using inappropriate noise characteristic is not a bias but rather sub-optimal statistical uncertainty. This however also implies again some implicit assumptions about the mixing matrix variability as we use its lower resolution estimates for the higher resolution maps. If this is justifiable then the map resolution used on this step could be even higher than that of the largest input beam and as high as the one for which this assumption holds. We discuss some of the aspects of this choice in the following noting here that in practice making the right choice can be rather difficult and objective-dependent. Indeed, it may well be that using the underpixelized, and oversmoothed component maps is the most robust way forward for cosmological parameter estimation, while higher resolution maps with more complex noise and potentially higher residuals can still be a useful source of information for other science goals,  e.g. Galactic science~\cite{hensley2022simons}.

All of the above issues can be sidestepped if we explicitly account for the beams in the component separation problem. 
Indeed, let us start from the data model in Eq.~(\ref{data_model_beams}) and introduce the target beam of the output component maps, hereafter denoted as $\mathbf{B}_\mathbf{final}$. Hereafter, we implicitly assume that the pixel size is set as dictated by the smallest beam out of the ones considered in the problem.
Algebraically, the final beam enters the data model as,
    \begin{eqnarray}
        \mathbf{d} & = & \mathbf{B}_\mathbf{true}^{\phantom{-1}}\,\mathbf{A} \,\mathbf{\hat{B}}^{-1}_\mathbf{final}\,\mathbf{\hat{B}}_\mathbf{final}^{\phantom{-1}}\,\mathbf{s}  + \mathbf{n}  \nonumber\\
& = & \mathbf{B}_\mathbf{true}^{\phantom{-1}}\,\mathbf{A} \,\mathbf{\hat{B}}^{-1}_\mathbf{final}\,\mathbf{s}^{\phantom{-1}}_\mathbf{final}  + \mathbf{n}, 
        \label{data_model_beam1}
    \end{eqnarray}
where $\mathbf{s}_\mathbf{final}$ are the beam-smoothed component templates. 
While any value of $\mathbf{B}^{\phantom{-1}}_\mathbf{final}$ can in principle be assumed, some values results in better numerical stability. We discuss this later on.
We note that this data model makes no assumptions about the commutation between (any of) the beam operators and the mixing matrix. In the following we refer to this case as \textit{BIC-Ncom}. 

In the cases where the commutation of the final beam and the mixing matrix can be assumed, i.e.,
\begin{equation}
    \mathbf{B}_\mathbf{final}^{-1} \, \mathbf{A} \; = \; \mathbf{A} \, \hat{\mathbf{B}}_\mathbf{final}^{-1}
     \label{eqn:BfinalAcommutation}
\end{equation}
the data model can be further simplified to read,
    \begin{eqnarray}
        \mathbf{d} & = & \mathbf{B}_\mathbf{true}^{\phantom{-1}}\,\mathbf{B}^{-1}_\mathbf{final}\,\mathbf{A} \, \mathbf{s}^{\phantom{-1}}_\mathbf{final}  + \mathbf{n}. 
        \label{data_model_beam-final}
    \end{eqnarray}
In the following we refer to this case as \textit{BIC-com}. 

While the validity of the commutation has to be evaluated for each experimental setup and spatial variability allowed for in the mixing matrix, see Section~\ref{sec:commutation} for more details, the advantage of this data model is that it permits a numerically more efficient implementation resulting in reduced computational cost with respect to the previous case. We also note that this data model is equivalent to the data model in Eq.~(\ref{preprocessing_data}), where the mixing matrix is kept in its original form but the noise covariance is made more complex, as opposed to the data model in Eq.~(\ref{data_model_beam-final}), where it is the mixing matrix which is modified. This is however only strictly true if the commutation relation in Eq.~(\ref{eqn:BfinalAcommutation}) is fulfilled.
We emphasize that whichever method is used the (de)convolution of the beam will always lead to some spurious effects whenever only partial sky is observed.  Indeed, for any pixel located within the beam size from the boundary any such operation is bound to fail. In the following we discuss the relevance of this effect in the context of the proposed method. 
We also note that the beam correction as discussed here is similar to the treatment used in some sampling based component separation codes, e.g., COMMANDER~\cite{eriksen2006cosmic}. These methods perform the component separation while characterizing simultaneously statistical properties of the components, such as their power spectra. Such applications can in principle avoid the boundary problem mentioned above as they estimate full sky samples conditioned on the observed, cut-sky data. The price to pay are the assumptions concerning the statistical properties of the sky components, such as Gaussianity and stationarity.

\section{Method}\label{sec:method}
\subsection{Framework}
We employ the two-step parametric component separation method, Eqs.~(\ref{eq:spec-like-def}) and~(\ref{eq:map-make-step}), ~\cite{stompor2009maximum},
assuming that the input data model is as given in Eq.~(\ref{data_model_beam1}) or in Eq.~(\ref{data_model_beam-final}).
The formalism therefore remains the same and the only change is that of the mixing matrix, $\mathbf{A}$, which is now replaced with its generalized version denoted as $\mathbf{\tilde{A}}$
which reads either,
\begin{equation}
    \mathbf{\tilde{A}} \equiv \mathbf{B}^{\phantom{-1}} _\mathbf{true}\, \mathbf{A} \, \mathbf{B}_\mathbf{final}^{-1},
     \label{eq:genMixingNonComm}
\end{equation}
for the BIC-Ncom data model or,
\begin{equation}
    \mathbf{\tilde{A}} \equiv \mathbf{B}^{\phantom{-1}} _\mathbf{true}\, \, \mathbf{B}_\mathbf{final}^{-1} \,\mathbf{A},
     \label{eq:genMixingComm}
\end{equation}
for the BIC-com data model.

The component separation problem, Eqs.~(\ref{eq:spec-like-def}) and~(\ref{eq:map-make-step}), is then solved by maximizing,
\begin{align} \label{sp_lik_tilde}
-2 \mathrm{ln} & \mathcal{L}_{sp} (\beta_i) 
     = \; \mathrm{const} \\
     & - \, (\mathbf{d}^T \mathbf{N}^{-1} \mathbf{\tilde{A}}) (\mathbf{\tilde{A}}^T \mathbf{N}^{-1} \mathbf{\tilde{A}})^{-1}(\mathbf{\tilde{A}}^T \mathbf{N}^{-1} \mathbf{d}). \nonumber
\end{align}
and estimating the component maps via,
\begin{equation} \label{Wd_tilde}
    \mathbf{s}_\mathbf{final} = (\mathbf{\tilde{A}}^T \mathbf{N}^{-1} \mathbf{\tilde{A}})^{-1} \mathbf{\tilde{A}}^T \mathbf{N}^{-1} \mathbf{d}.
\end{equation}
While conceptually simple, the need for accommodating the generalized mixing matrices calls for more involved numerical algorithms needed for the implementation of the method. We describe them in the following section.

\subsection{Implementation}\label{implementation}

\subsubsection{Algebraic operations}
The algebraic operations involved in these two steps can be in general performed either in harmonic or pixel domain, each having its own advantages.
In the harmonic domain applying the beams is more straightforward, at least for the full sky coverage and axially symmetric beams as each beam operator $\mathbf{B}$ is represented by a diagonal matrix.
In the pixel domain instead we can easily account for the spatial variability of the foregrounds by allowing the mixing matrix to be pixel-dependent.
The noise inhomogeneities are easier to treat in pixel domain, as is the general case of the non-sky-stationary correlated noise, the computational load needed for this latter case notwithstanding. 
The harmonic domain allows for easy treatment of the sky-stationary noise, which in some applications can be a sufficient approximation. 
Here, we choose an alternative approach which tries to capitalize on advantages of both these domains.
Therefore, each matrix in Eqs.~(\ref{sp_lik_tilde}) and~(\ref{Wd_tilde}) is applied in its most suitable domain and the necessary back and forth transforms between the pixel to harmonic domain are performed using spherical harmonic transforms (SHTs).
Specifically, in our implementation the noise correlations matrices, as well as the component mixing matrix, $\mathbf{A}$, are applied in the pixel domain.
This allows us to deal rigorously with the cut sky, inhomogeneous noise, and the spatial variability of the foregrounds effects that are naturally living in the pixel domain. 
A general (non-sky-stationary) correlated noise should also be treated preferentially in pixel domain though it could be computationally rather involved as it may in general require map-making type operations within the component separation code. We leave such an extension for future work.

A sky-stationary correlated noise can instead be accounted for most straightforwardly in the harmonic domain and treated together with the beams, as the beam operators are applied in the harmonic domain. 
Given that we do not have typically access to the full sky we can not recover actual harmonic representation of the sky maps. 
We therefore can deal instead only with the pseudo harmonic coefficients, which are formally a convolution of the observed sky mask and the true harmonic coefficients.
While it is an approximation, it is usually a very good one as the modes of interest, i.e., those affected by the beam smoothing, are the ones much shorter than the observed patch size. 
In pixel domain, for well-behaved beams, we expect the failures to be well localized and limited to a narrow strip of pixels along the patch boundary and that they quickly vanish when we move away from it towards the patch center. 

There are three basic operations we need to be able to perform in order to calculate the likelihood and component signal estimates in Eqs.~(\ref{sp_lik_tilde}) and~(\ref{Wd_tilde}). These are:
\begin{description}[font=$\bullet$~\normalfont]
\item[\textbf{noise weighting of frequency domain maps}, $\mathbf{N}^{-1}\,\mathbf{m}$] {in the current implementation this is performed directly in the pixel domain owing to the assumption of diagonal noise covariances.
If needed the sky-stationary, correlated noise component can be easily added as appropriate  noise weighting applied in the harmonic domain.}
\item[\textbf{generalized component mixing}, $\mathbf{\tilde{A}}\, \mathbf{s}$] {this is performed as a sequence of operations with the operators applied to pixel domain objects from right to left. All the beam operations pass through  harmonic domain, while the actual mixing of the components, i.e., $\mathbf{A}$, is applied directly in the pixel domain. This requires back and forth transformations between the pixel and harmonic domains. The number of the necessary transformations depends on the adopted data model, with the generalized mixing matrix in Eq.~(\ref{eq:genMixingNonComm}) requiring twice as many of those as the mixing matrix in Eq.~(\ref{eq:genMixingComm}). 
}
\item[\textbf{transposed generalized component mixing}, $\mathbf{\tilde{A}}^T\, \mathbf{m}$] {this operation is also implemented as a sequence of operations, analogous to the case above but performed in the reverse order.}
\end{description}

These operations suffice for the computation of $\mathbf{\tilde{A}}^T \mathbf{N}^{-1} \mathbf{d}$. The computation of the full covariance matrix, $\mathbf{\tilde{A}}^T \mathbf{N}^{-1} \mathbf{\tilde{A}}$ is more involved as in the case at hand it does not have simple, block diagonal structure with the blocks corresponding to different observed sky pixels. Thus in general the construction and the inversion of the full pixel-domain covariance is therefore only possible for the lowest resolution cases with a limited number of sky pixels.
These problems can be overcome in a specific case of a pixel-independent mixing matrix, $\mathbf{A}$, and homogeneous noise. We can then apply both the mixing matrix and the noise covariance matrix directly in the harmonic domain and represent all the map objects via their spherical harmonic representations. Then matrix $\mathbf{\tilde{A}}^T \mathbf{N}^{-1} \mathbf{\tilde{A}}$, which would then operate on harmonic coefficients, would be block-diagonal and thus simply invertible block-by-block.

In cases of more general component mixing or noise patterns, such approaches are not applicable, and instead we develop an iterative solver to compute directly a product of the inverse noise covariance and a pixel-domain map-like vector.


In our approach we use a preconditioned conjugate gradient (PCG) solver to solve the system,
\begin{equation}
    \mathbf{\tilde{A}}^T \mathbf{N}^{-1} \mathbf{\tilde{A}} \mathbf{x} = \mathbf{\tilde{A}}^T \mathbf{N}^{-1} \mathbf{d},
    \label{eq:compsep_step1}
\end{equation}
obtaining the current component map estimates, $\mathbf{x}$, which we then insert back in the spectral likelihood,
\begin{equation}
    -2 \mathrm{ln} \mathcal{L}_{sp} (\beta_i) 
    = \mathrm{const} - (\mathbf{d}^T \mathbf{N}^{-1} \mathbf{\tilde{A}}) \mathbf{x}.
\end{equation}
The implementation relies on the PCG solver of \texttt{scipy}~\cite{2020SciPy-NMeth}.
The operations discussed earlier are sufficient for the implementation of all the computations required by the PCG solver. To accelerate it, we employ a preconditioner constructed, and applied, in pixel domain.
It looks like the system matrix but with the actual beam free mixing matrix $\mathbf{A}$, with the spectral parameters fixed at their current estimate: $\mathbf{A}^{T} \mathbf{N}^{-1} \mathbf{A}$.
This preconditioner is able to  account for the spatial variability of the foregrounds, but not the beams.
Consequently this PCG solver converges rapidly to a solution at the largest angular scales, but needs many more iterations to converge the smaller angular scales, where the beams impact is more important and the preconditioner inefficient.

We have also explored an alternative preconditioner which is expected to perform well if the mixing matrix variability across the sky is sufficiently small.  
The preconditioner is constructed as the inverse noise covariance as described above in the case of homogeneous noise and pixel-independent mixing matrix. It is thus built and applied directly in the harmonic domain where it has a block-diagonal structure and, by construction, can account nearly strictly on the beam effects. Due to sky variability of the mixing matrices considered in the examples studied in this work, this preconditioner was found to not be competitive with the one described earlier. Nevertheless, some hybrid constructions retaining the advantages of both could outperform any of the two. As could some more generic preconditioners, i.e., such as those based on two-level constructions~\cite{szydlarski2014accelerating, el2022mappraiser}. We leave such investigations to future work and use the first preconditioner as proposed here throughout this work.

Our implementation is based on the \texttt{fgbuster} code~\cite{fgbuster}, but it accounts explicitly for the beams in the mixing matrices and uses the PCG solver to evaluate the spectral likelihood.
The effective mixing matrix $\mathbf{\tilde{A}}$ is implemented as an operator, which applies it to any arbitrary pixel-domain vector without however ever constructing it. In the cases where the sky-stationary noise correlations are present, and described by the noise power spectrum, $\boldsymbol{\mathcal{N}}_\ell$,  we first modify the beam function, $\mathbf{b}_{\mathbf{true}, \ell} \rightarrow \mathbf{b}_{\mathbf{true}, \ell}\,\boldsymbol{\mathcal{N}}_\ell^{-\alpha}$, $\alpha=0.5$ or $=1$ for the left, respectively right, hand side operators in Eq.~(\ref{eq:compsep_step1}) and then call the operator as usual. 
This operator is then used in all the generalized spectral likelihood calculations as well as in the PCG solver. 
The code is released as part of the \texttt{fgbuster} package, at \cite{fgbuster-beams}.

\subsubsection{Likelihood optimization}
The extended \texttt{fgbuster} package includes also all the functionality necessary for the maximization of the spectra likelihood which constitutes the first step of the component separation approach. It has been implemented and validated as part of the present work. Consequently, though this functionality has no direct bearing on the results presented hereafter, for completeness, we provide below a brief description of details of the numerical implementation.

The likelihood maximization is performed with help of the \texttt{optimize.minimize} module of the \texttt{scipy} package~\cite{2020SciPy-NMeth}.
This package offers various methods to reach the convergence.
As long as we fit for a limited number of spectral parameters we were able to reach the convergence with the Broyden–Fletcher–Goldfarb–Shanno (BFGS) algorithm computing the gradient of the spectral likelihood numerically.
Still, this implementation was quite slow to converge, due to the fact that each evaluation of the spectral likelihood takes order of minutes to converge even for lower resolution maps, mostly driven by the number of PCG iterations required and the SHTs cost when considering higher resolution cases.
To minimize the number of the spectral likelihood evaluations,
we have implemented an analytical expression for the gradient of the spectral likelihood, given by,
\begin{align}
    -2 \; \mathbf{\nabla}_{\beta} \mathrm{ln} \mathcal{L}_{sp} = -2 \mathbf{d}^T \mathbf{N}^{-1} \mathbf{\tilde{A}} \left( \mathbf{\tilde{A}}^T \mathbf{N}^{-1} \mathbf{\tilde{A}} \right) ^{-1} \mathbf{\tilde{A}},_{\beta} ^T  \mathbf{P} \mathbf{d}, \ 
\end{align}
where,
\begin{equation}
    \mathbf{P} \equiv \mathbf{N}^{-1} - \mathbf{N}^{-1} \mathbf{\tilde{A}} \left( \mathbf{\tilde{A}}^T \mathbf{N}^{-1} \mathbf{\tilde{A}} \right) ^{-1} \mathbf{\tilde{A}}^T \mathbf{N}^{-1}.
\end{equation}
\\
We note that the only additional operations required in this computation involve an application of the derivative of the generalized mixing matrix with respect to the spectral parameters to a frequency map vector. 
This is straightforward and computationally cheap as it can be done without any additional SHTs, for instance, via generalizing the general mixing matrix operator so it performs simultaneously the product of the generalized mixing matrix and its derivative by an arbitrary (but common) pixel-domain vector.
All the other operations, in particular those included in the PCG iterations, have to be computed anyway for the spectral likelihood evaluations and reused here.
This allows us to use more efficient optimization algorithms, which require fewer computations of the likelihood function.
This optimization is particularly important when we fit for many spectral parameters as in the cases with spatial variable foreground scalings. 

\subsubsection{Numerical efficiency}
As mentioned in the previous paragraph, the computational time of a single likelihood evaluation is essentially proportional to the number of iterations required for the PCG solver to converge. This emphasizes the need for very efficient preconditioners. 
In the next section we discuss the performance of the preconditioner proposed here.

For a given PCG iteration, the computation time increases with the resolution of the maps, as it is dominated by the time needed to compute the SHTs.
These need to be performed every time we apply the generalized mixing matrix or its transpose. 
At present this is done with help of the \texttt{healpy} package~\cite{Zonca2019, 2005ApJ...622..759G}, however, we currently explore a potential use of a \texttt{JAX}-based package (and GPUs). We anticipate that this will significantly alleviate the computational limitations of the current code and will be made available in the forthcoming release. All the results shown in this work have been produced with \texttt{healpy}-based code.

\section{Results}\label{sec:results}
We present in this section some results of the method presented above with the goal of comparing the validity and performances of the different data models, BIC-Ncom and BIC-com, among themselves and with the reference CRA, in realistic experimental scenarios.
We consider in the following only beams given by a main lobe and with circular Gaussian profile, however any axially symmetric beam profile could be
assumed instead if needed.

The impact of the beam treatment in the component separation can be expected to depend on the specifics of the instrumental setup assumed, in particular on the range of resolution of the input frequency bands of a given instrument: the broader this range the more significant the impact will be.
We therefore show the results for two different experimental setups, spanning the range of possible choices.
We emphasize that the cases studied hereafter are highly idealized as our goal is to validate and evaluate the new beam treatment and not to explore yet all possible complications. We note, however, that many of the relevant effects, such as bandpasses, beam chromaticity, etc., which without a doubt will be important in actual applications, e.g.~\cite{leloup2024impact, dachlythra2024simons}, could be included in our formalism rather straightforwardly. We leave that to future work.

\subsection{Large sky fractions}
The first setup we consider is that of a full sky mission, with resolution of the input frequency bands going from $17.9'$ to $70.5'$, as it would be the case for the $LiteBIRD$ instrument~\cite{litebird2023probing}.
In Table~\ref{tab:LB-like} we list all the specifics of this setup, which we will refer to as $LiteBIRD$-like setup.
\begin{table}[h]
\centering
\begin{tabular}{|c|c|c|c|}
\hline
\multicolumn{1}{|c|}{$\nu$} & sensitivity & FWHM \\
\multicolumn{1}{|c|}{[GHz]} & $[\mu \mathrm{Karcmin}]$ & [arcmin] \\
\hline
40.0 & 37.42 & 70.5 \\
50.0 & 33.46 & 58.5 \\
60.0 & 21.31 & 51.1 \\
68.0 & 16.87 & 47.1 \\
78.0 & 12.07 & 43.8 \\
89.0 & 11.30 & 41.5 \\
100.0 & 6.56 & 37.8 \\
119.0 & 4.58 & 33.6 \\
140.0 & 4.79 & 30.8 \\
166.0 & 5.57 & 28.9 \\
195.0 & 5.85 & 28.6 \\
235.0 & 10.79 & 24.7 \\
280.0 & 13.8 & 22.5 \\
337.0 & 21.95 & 20.9 \\
402.0 & 47.45 & 17.9 \\
\hline
\end{tabular}
\caption{Frequency, sensitivity and FWHM of Gaussian beams for the $LiteBIRD$-like setup. Taken from~\cite{litebird2023probing}.}
\label{tab:LB-like}
\end{table}

The frequency maps are simulated at \texttt{nside}~=~512, and the component separation is performed on the full sky, with a galactic plane mask at $f_{\rm{sky}}=60\%$ from~\cite{HFI_Mask} applied on the recovered components when necessary.
To create the simulations we perform the following steps: first the sky signal is simulated through the \texttt{PySM}~\cite{thorne2017python, zonca2021python} package, the true beams of Table~\ref{tab:LB-like} are then applied, finally the noise is simulated as a Gaussian realisation and added to the convolved sky signal.
As sky model we take the \texttt{PySM} $c1$ model as a CMB realisation with lensing and no primordial $B$ modes, to which we add thermal dust and synchrotron emissions.
In this work we assume the scaling of the CMB to be known, hence the parametric component separation results do not depend on the input CMB, and the CMB residuals, due to misestimated foreground contributions, do not depend on the specific CMB realization.
For the foreground components instead we consider in the following two options, first the \texttt{PySM} model $d0s0$, which simulates the dust frequency scaling with a modified black body and the synchrotron one with a power law, assuming as spectral parameters the spectral index of the modified black body and its temperature, and the spectral index of the power law, respectively with values, $\beta_{\text{dust}}=1.54$, $\text{T}_{\text{dust}}=20$~K and $\beta_{\text{synch}}=-3$ across the full sky.
We then consider the $d1s1$ model, again given by a modified black body and a power law but with spectral parameter values varying across the sky, with most of the variations along the Galactic plane.
This latter model is adequate for our scope of exploring the effect of dealing with the beams in a scenario with spatially varying scaling laws, and a more complex sky model with a higher level of spatial variability would still suffer the complications of those in an enhanced way with respect to what presented below, according to its level of spatial variability.

\subsubsection{Without spatial variability}
\begin{figure*}
    \centering
    \includegraphics[width=2\columnwidth]{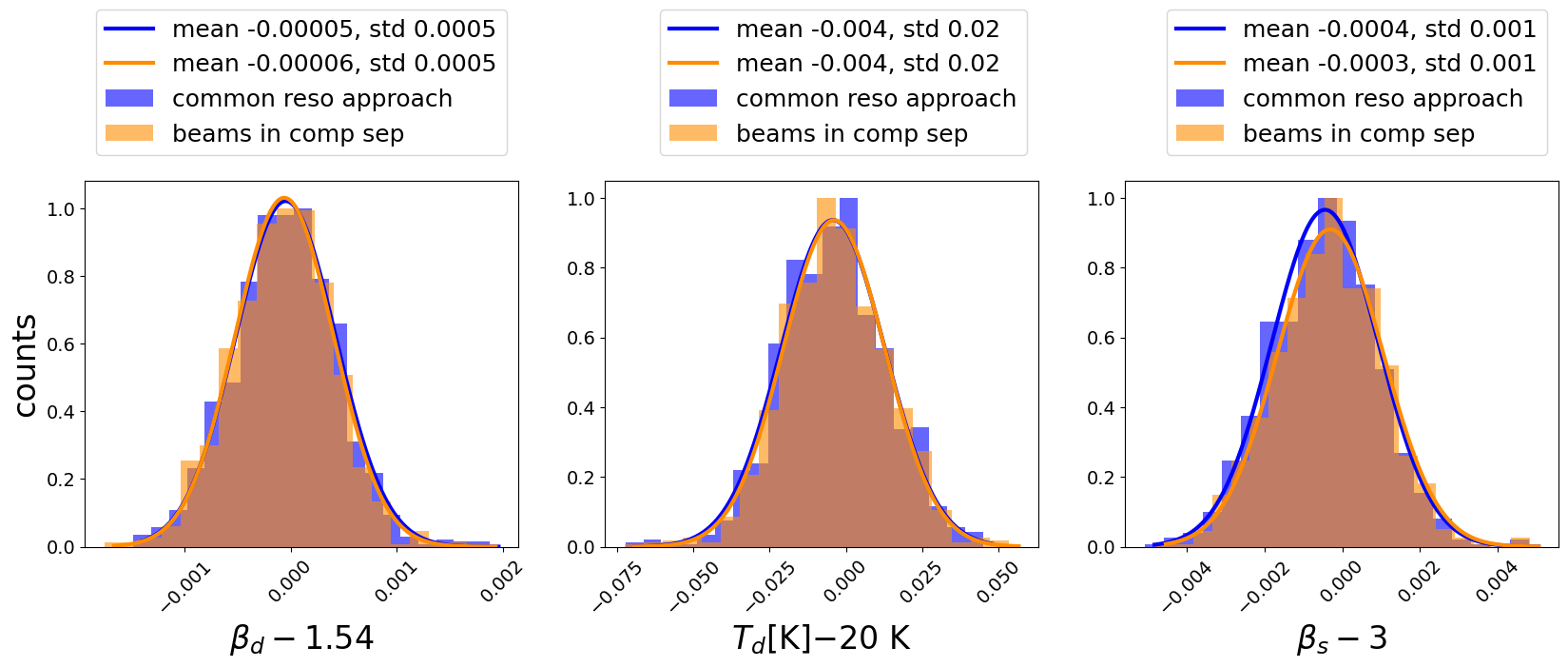}
    \caption{Recovered spectral parameters from component separation on $1,000$ simulations with different noise realizations. The horizontal axis give the difference with the true spectral parameter values. \textit{In blue}: CRA, dealing with the beams before the component separation with $\mathbf{B}^{\phantom{-1}}_\mathbf{final}=70.5'$ and downgrading to \texttt{nside}~=~32; \textit{in orange}: BIC,
    dealing with the beams in the component separation.
    The two distributions are both peaked on the true spectral parameter values and have very close width, given the number of samples they are statistically consistent.}
    \label{fig:improvment_on_betas}
\end{figure*}

We apply here the proposed BIC approach to the simulations without spatial variability of the foreground properties and compare it with the CRA. As the assumed scaling laws are pixel independent the beam and mixing operators commute. The expected differences in performance of different methods is therefore due to treatment of the map resolution and the noise.
For the CRA we perform the preprocessing by convolving each input frequency map with the effective beam $\mathbf{B}^{\phantom{-1}}_\mathbf{final} \mathbf{B}^{-1}_\mathbf{true}$, where the first term is a Gaussian beam with full width half maximum (FWHM) given by the largest instrumental beam, $70.5'$.
In addition, for the sake of the first step of the component separation, we downgrade to the pixel resolution of \texttt{nside}~=~32, in order to suppress the noise correlations the additional smoothing unavoidably introduces.
The value of $\mathbf{B}^{\phantom{-1}}_\mathbf{final}$ in general is chosen in a way that the enhancement of the noise power due to the true beam deconvolution, $\mathbf{B}_\mathbf{true}^{-1}$, is suppressed. As this is followed up by the underpixelization of the maps, there is some freedom in selecting the final beam size.
Downgrading to such a low \texttt{nside} dramatically decreases the angular scales contributing to determining the spectral parameters. 
In the case of the pixel-independent foreground properties this may only lead to some loss of precision. While we could try using larger \texttt{nside} to compensate for this, this poses a potential risk due to a bias ensued by the remaining correlations in the noise, which are not easily incorporated in the noise model. While there may be some potential trade-off to explore, we use, hereafter, \texttt{nside}~=~32 as the resolution of the downgraded map.
The spectral parameter estimates recovered then on the first step of the component separation are subsequently used to produce the component maps on the second step.
In this latter case, we can in principle keep the original pixel resolution of \texttt{nside}~=~512 and choose the final beam size accordingly. 
As the second step is linear, this choice affects the noise level of the produced maps, but does not result in biases, and the beam size can be then optimized accordingly.
For simplicity, unless mentioned otherwise, we adapt here the value of $70.5'$ (FWHM) for $\mathbf{B}^{\phantom{-1}}_\mathbf{final}$, corresponding the largest experimental beam, to efficiently compensate for $\mathbf{B}_\mathbf{true}^{-1}$.
As far as the BIC method proposed here is concerned, given that the beam and mixing operators commute the data models in Eqs.~(\ref{data_model_beam1}) and~(\ref{data_model_beam-final}) are equivalent, hence we do not need to distinguish between BIC-Ncom and BIC-com in this section. 
Moreover, as the noise covariance is also pixel independent we can perform all the operations in the harmonic domain (including the spectral likelihood minimization), and come back in the pixel domain only once we have the recovered components. 

This simplifies the comparison as we have only two methods to compare, the reference, CRA and one of the variants of the beam-treating approach, BIC. We first compare the estimation of spectral parameters. We do so by simulating $1,000$ input data sets with different instrumental noise realizations. The histograms are shown in Figure~\ref{fig:improvment_on_betas}. We see that both methods perform similarly. This may seem surprising given the fact that the BIC method should be able to exploit all the small scale information not accessible to the oversmoothed CRA. While in principle this is indeed so, we note that due to the redness of the power spectra of all non-CMB components the information concerning their spectral parameters residing at small-scales is essentially negligible.
\begin{figure}
    \centering
    \includegraphics[width=\columnwidth]{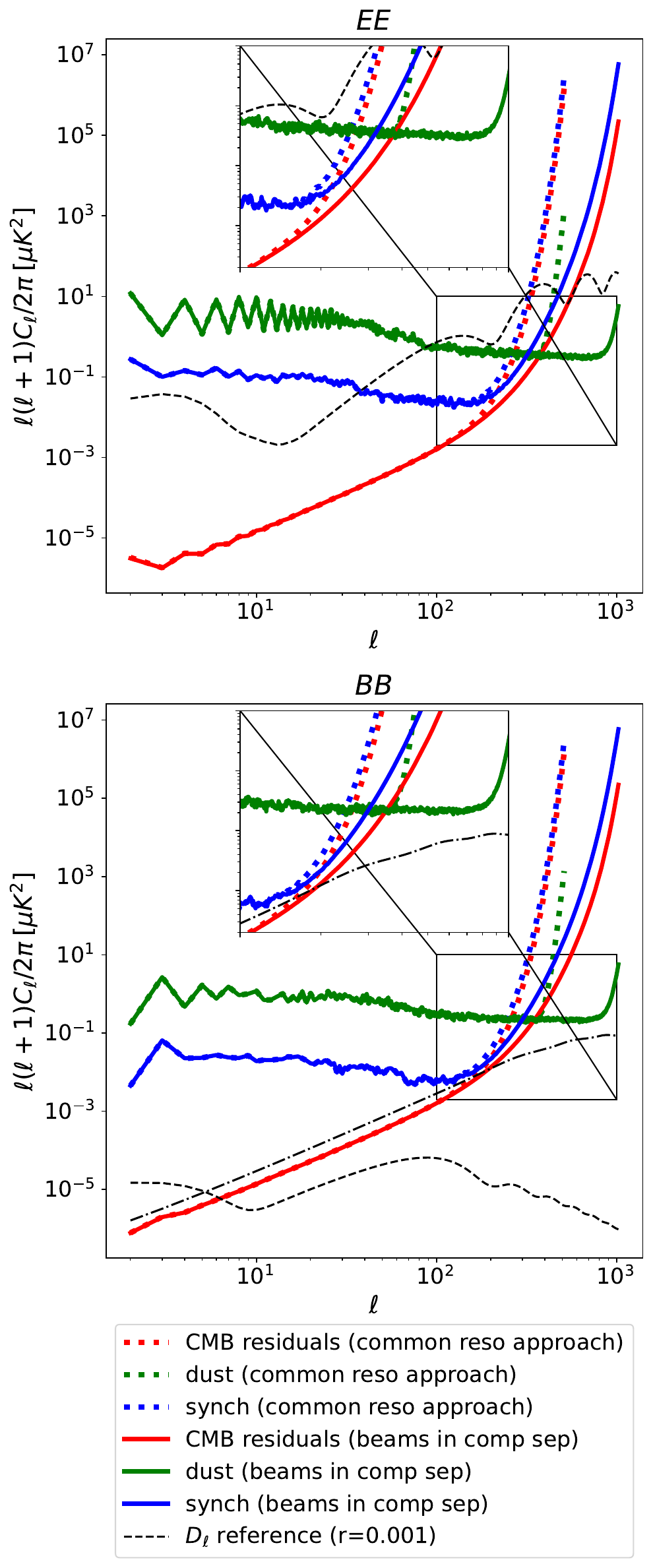}
    \caption{Average of the recovered foreground components and CMB residuals for $1,000$ noise realizations, for the $LiteBIRD$-like setup with input full sky \textit{c1d0s0}. The components are shown at the frequency of 100~GHz.
    With the proposed approach dealing with the beams in the component separation, BIC, we can access smaller angular scales as compared to the approach dealing with the beams before the component separation, CRA.}
    \label{fig:plot1}
\end{figure}
Instead the differences can be seen on the second step of the component separation, when the component maps are estimated.
As shown in Figure~\ref{fig:plot1} the impact of accounting for the beams in the component separation can be indeed significant.
Given the very simple foreground models assumed in this case, the CMB residuals are dominated by the instrumental noise after component separation.
We point out that the different recovered components start to be dominated by the noise at different scales: at about $\ell=150$ for synchrotron, at about $\ell=350$ for dust in the CRA and $\ell=800$ for dust in the BIC approach.

\begin{figure}
    \centering
    \includegraphics[width=\columnwidth]{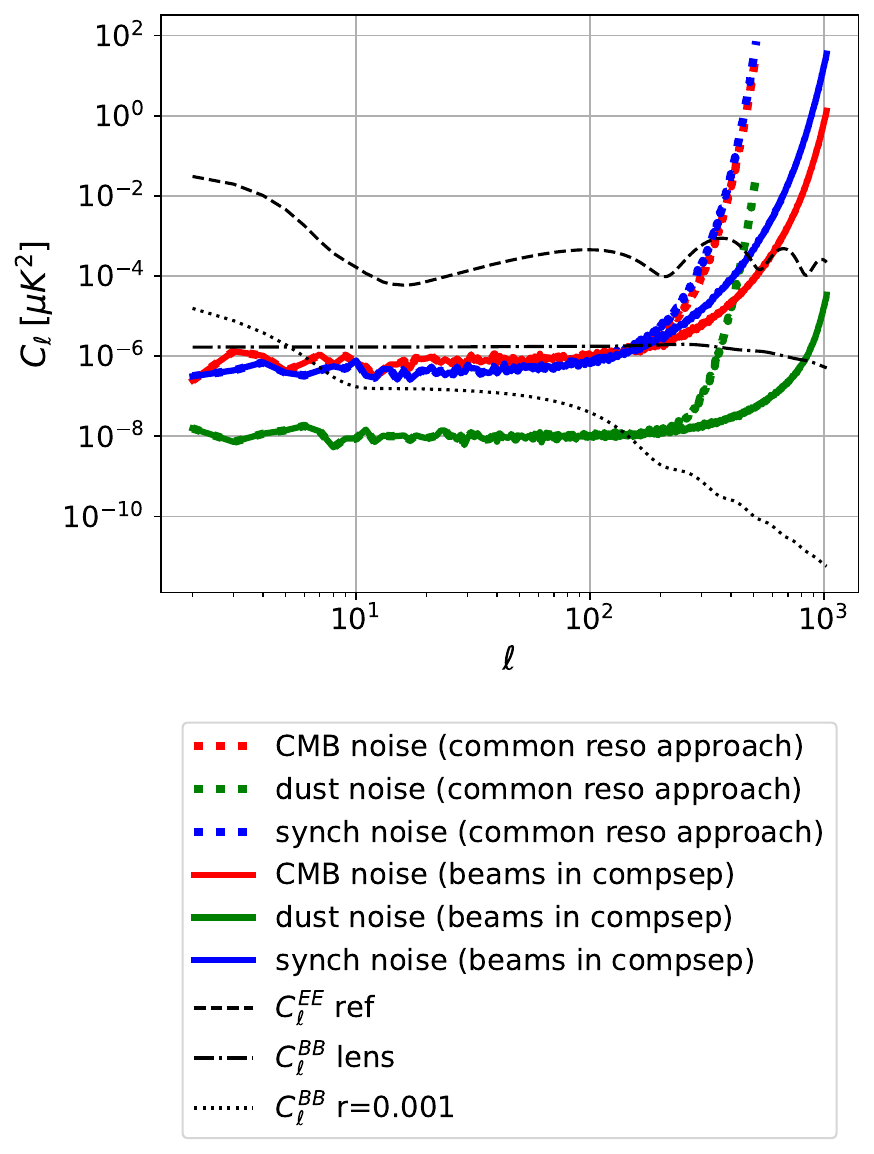}
    \caption{Noise spectra after component separation for a given simulation, for the two component separation approaches CRA and BIC.}
    \label{fig:noise_spectra}
\end{figure}

\begin{figure*}
    \centering
    \includegraphics[width=2\columnwidth]{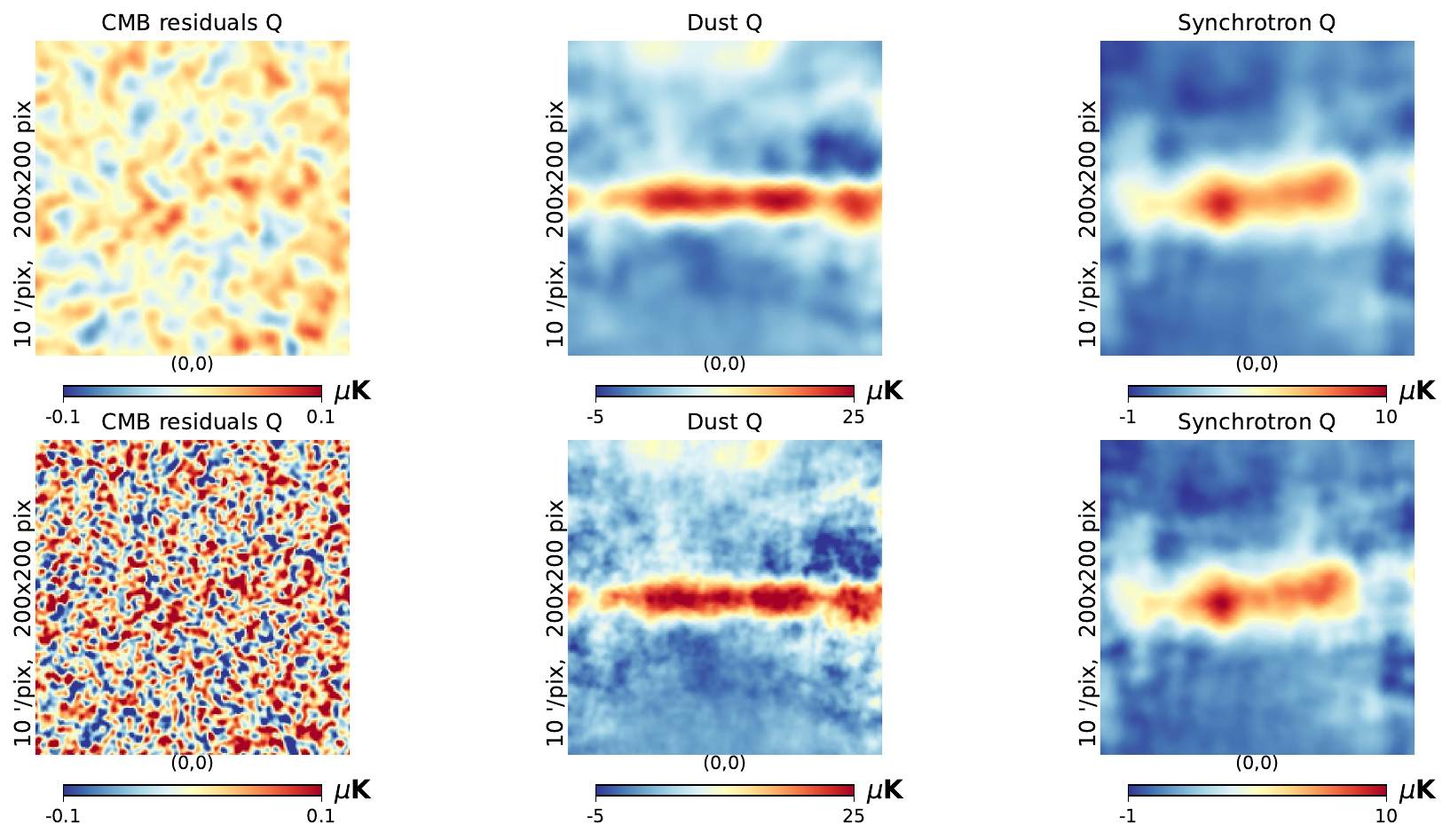}
    \caption{$LiteBIRD$-like experiment, input sky \textit{c1d0s0} full sky, component separation with CRA, first row, versus BIC approach,
    second row. 
    We show a patch of the sky for the recovered CMB residuals, thermal dust (at 100~GHz) and synchrotron (at 100~GHz) maps of about $30^\circ \times 30^\circ$ centered on the galactic plane. We plot the Q Stokes parameter.}
    \label{fig:patch_gal_plane}
\end{figure*}

\begin{figure*}
    \centering
    \includegraphics[width=2\columnwidth]{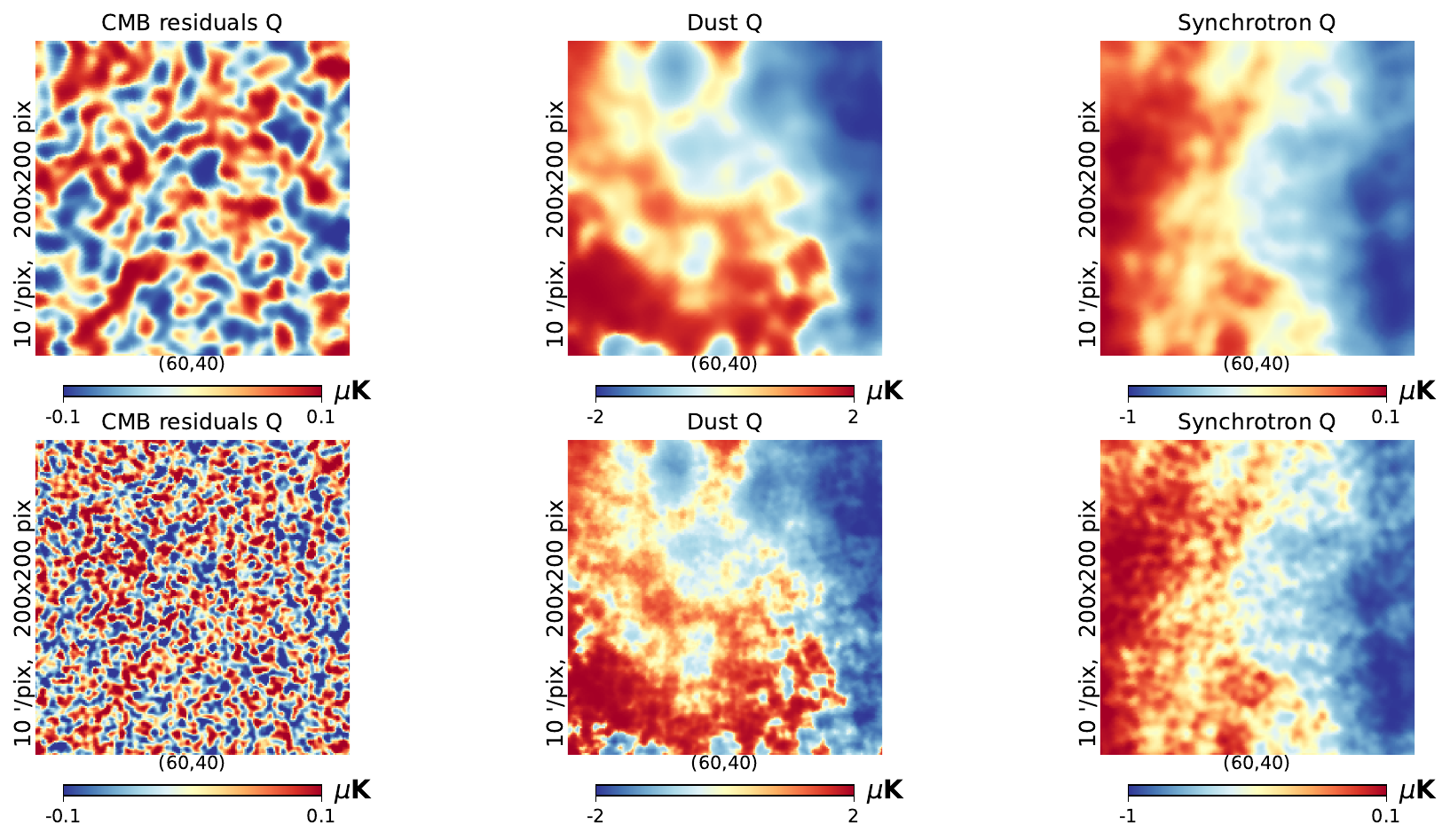}
    \caption{$LiteBIRD$-like experiment, input sky \textit{c1d0s0} full sky, component separation with CRA, first row, versus BIC approach, second row.
    We show a patch of the sky for the recovered CMB residuals, thermal dust (at 100~GHz) and synchrotron (at 100~GHz) maps of about $30^\circ \times 30^\circ$ centered in a clean region of the sky at longitude~=~$60^\circ$ and latitude~=~$40^\circ$.
    We plot the Q Stokes parameter.}
    \label{fig:patch_higher_lat}
\end{figure*}

The noise spectra after component separation are shown in Figure~\ref{fig:noise_spectra}.
In addition, the different recovered components have the noise exploding at different scales, which depends on the beam convolution remaining in the noise after component separation and which is given by a combination of all the $\mathbf{B}_\mathbf{true}^{-1}$, which varies for the different components, but is steeper in the CRA then in the BIC approach.
The gain in resolution in the latter approach is thus due to fact that the noise starts to dominate later and when it does it goes up in a less steep way.
The wider range of accessible scales with this method is clearly visible in the plot.
Also, the gain in resolution is more important for the dust component, than the synchrotron component.
This is because the additional resolution that we can now exploit comes from the higher frequency bands with smaller beam FWHM, and thus affecting mainly the dust component.
With the proposed approach about a factor 1.5 is gained in the recovered CMB residuals before the noise stops on being white and more than a factor two for the dust component.
We show a patch of the maps of the recovered CMB residuals for one realization of the sky, thermal dust emission and synchrotron emission on the Galactic plane in Figure~\ref{fig:patch_gal_plane} and in a cleaner region of the sky in Figure~\ref{fig:patch_higher_lat}.
In both cases the improvement in resolution is visible in the BIC approach, in particular for the dust component.
The maps shown here are smoothed with a Gaussian beam with the width as recovered from the noise power spectra shown in Figure~\ref{fig:noise_spectra} suppressing the excess noise on the smallest scales and defining the resolution. While better resolution is in principle possible it will come at the cost of significant noise excess.
The main goal of the proposed procedure is to facilitate access to the information on the smaller angular scales, which is perhaps the most important for Galactic science goals or for science using the Sunyaev Zel’dovich effects~\cite{gorce2022retrieving, douspis2024small}.
In terms of constraining cosmological parameters the most significant impact is expected for parameters relevant for small scales. 
Indeed with this particular setup we do not get a significant improvement on the uncertainty on $r$.
A mild improvement could instead be reached with other experimental setups, with a broader range of resolution of the input frequency bands or for experiments aiming at constraining $r$ from medium scales, which is for example the case of $SO$ or any other ground based experiment for which accessing the reionization peak in the $B$-mode power spectrum is excluded.

\subsubsection{With spatial variability}
For the foreground simulations we now consider the \texttt{PySM} $d1s1$ model with spectral parameter variations across the sky.
To have low CMB residuals from the component separation we are now obliged to account for the spatial variability of the mixing matrix.

\begin{figure}
    \centering
    \includegraphics[width=\columnwidth]{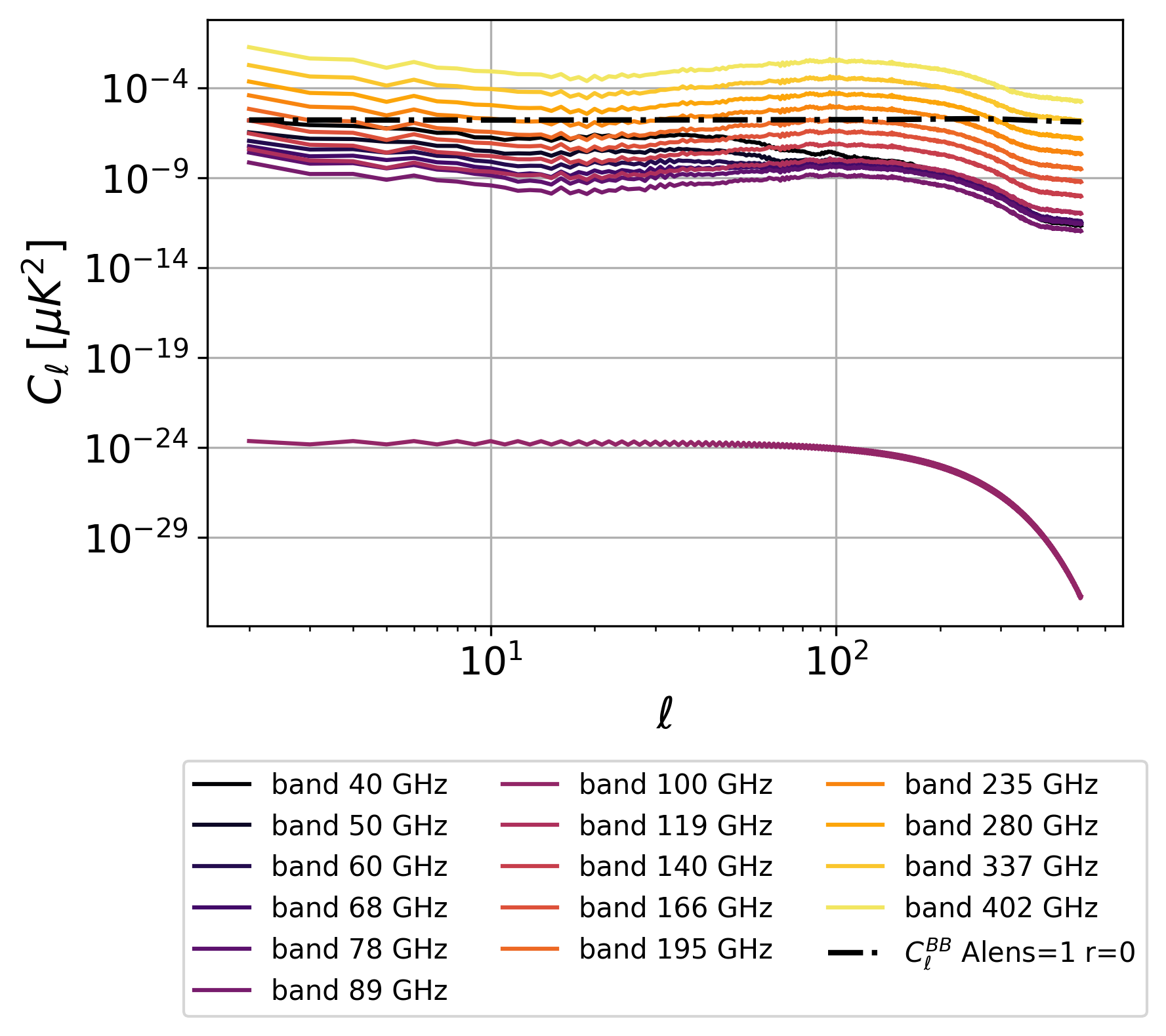}
    \caption{Commutation of $\mathbf{A}$ and $\mathbf{B}^{\phantom{-1}}_\mathbf{final}$ in the full-sky, true spectral parameter case, as it would occur in the CRA. We use maps at \texttt{nside}~=~512 and FWHM of $\mathbf{B}^{\phantom{-1}}_\mathbf{final}=70.5'$. The commutation introduces a non-negligible error for all the bands except that of $100$~GHz at which the mixing matrix is normalized and thus spatially invariant. The lensing $B$-mode curve is plotted for reference.}
    \label{fig:ABs-BAs_Cls}
\end{figure}

\begin{figure}
    \centering
    \includegraphics[width=\columnwidth]{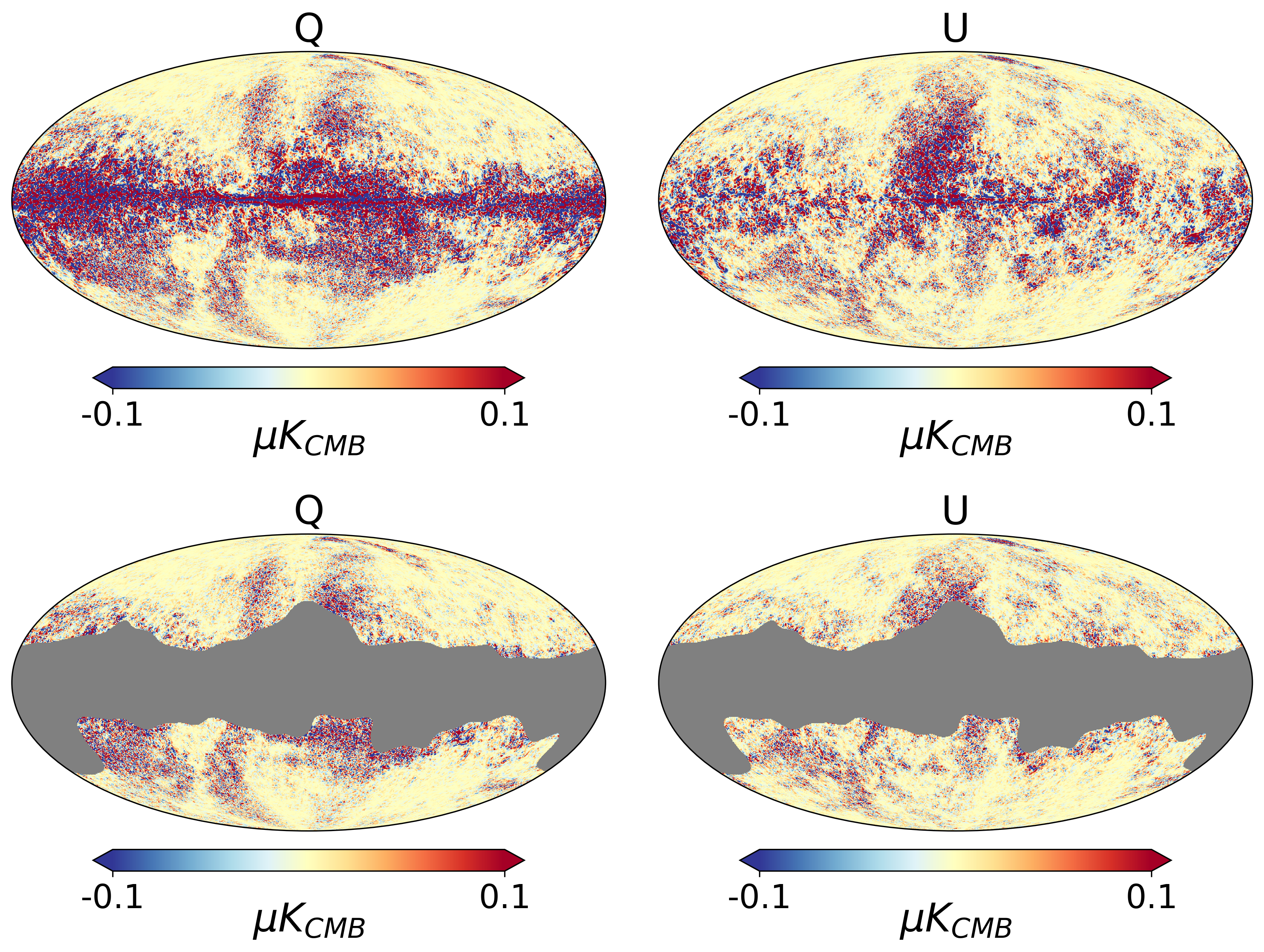}
    \caption{Commutation of $\mathbf{A}$ and $\mathbf{B}^{\phantom{-1}}_\mathbf{final}$ for the 235~GHz band. The plot shows $\mathbf{ABs-BAs}$ with and without masking the Galactic plane. The mask used is the $f_{\rm{sky}}=60\%$ mask of \cite{HFI_Mask}.
    We note that the values depend on the mixing matrix normalization frequency, set here to be $100$~GHz, at which the mixing displays no spatial variability.
    Bands closer to 100~GHz will thus have smaller amplitudes of these residuals, while these will be larger for bands further away from 100~GHz. See Fig.~\ref{fig:ABs-BAs_Cls}.}
    \label{fig:ABs-BAs_maps}
\end{figure}

\begin{figure}
    \centering
    \includegraphics[width=\columnwidth]{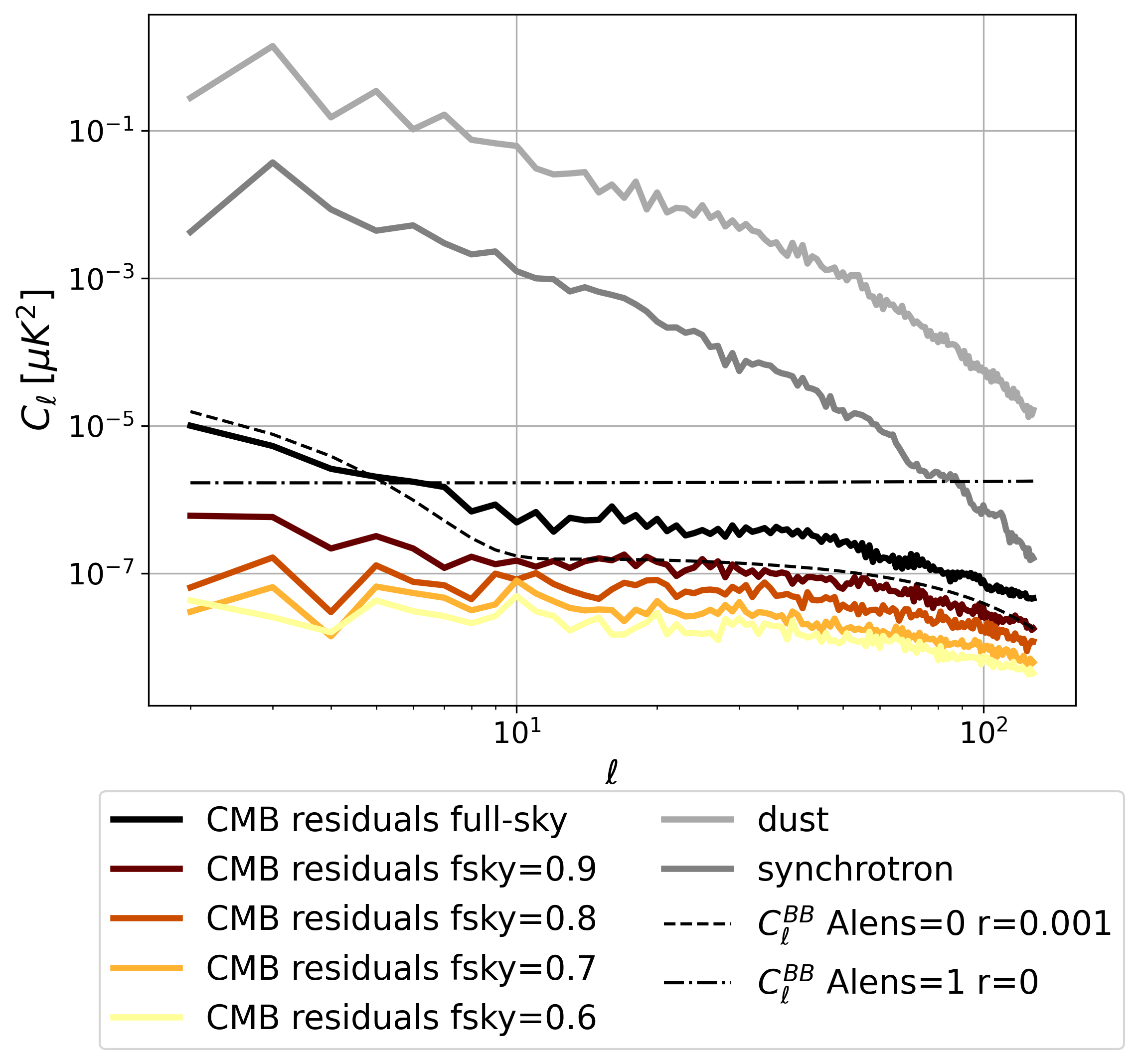}
    \caption{$BB$ CMB residuals due to the beam in the CRA, arising both due to the non-commutation of the mixing matrix with the beam operator and the smoothing of the original maps with the beam. The residual is comparable with the targeted signal for the full-sky case, and is progressively reduced by masking the Galactic plane where the spectral parameters present the most spatial variability.}
    \label{fig:commutation_ABeff_compsep}
\end{figure}

We first explore the effect of performing the commutation of the mixing matrix and the beam operator, as in Eq.~(\ref{eqn:BAcommutation}), in the data preprocessing of the CRA.
We do this by computing the difference,
\begin{equation}
    \mathbf{B}_\mathbf{final} \mathbf{A} \mathbf{s} - \mathbf{A} \mathbf{B}_\mathbf{final} \mathbf{s},
\end{equation}
where $\mathbf{A}$ was computed using the true $d1s1$ spectral parameter values,
$\mathbf{s}$ are the component maps at \texttt{nside}~=~512, and $\mathbf{B}_{\mathbf{final}}$ a Gaussian beam with FWHM $70.5'$.
We find this difference to be non-negligible for all the frequency bands, except for the one at which we have normalized the mixing matrix, which is spatially constant in that band, Figure~\ref{fig:ABs-BAs_Cls}.
The difference is particularly noticeable along the Galactic plane, as this corresponds to the sky region where the spectral parameters considered are varying the most, Figure~\ref{fig:ABs-BAs_maps}.
While at face value this difference seems clearly important, of real interest is how it may affect the recovered components, and how much this depends on whether the Galactic plane is masked or not.
We thus compute the residuals that we would have because of the beams in the case of noiseless input simulations and with the mixing matrix with one patch per pixel, at \texttt{nside}~=~64, and with $\mathbf{B}^{\phantom{-1}}_\mathbf{final} = 70.5'$.
Doing so we ensure that there is no bias due to the choice of the scaling laws in the mixing matrix, and the residuals that we find are entirely due to the effect of the beam commutation, the same setup without beams would indeed give numerically zero residuals.
Instead we find that the residuals can be comparable with the target signal in the full-sky case, while they decrease in the case where we mask the Galactic plane, Figure~\ref{fig:commutation_ABeff_compsep}.
While the discussion of this section is generally valid both for $E$ and $B$ modes, we focus here on the $B$ modes and show those in the figures.

\bigskip
To account for the spatial variability of the mixing matrix in the following we consider a simplified version of what was done in~\cite{litebird2023probing}, where they allow for the spatial variability of the mixing matrix for each spectral parameter on different sky pixels, corresponding to different values of \texttt{nside}.
As our goal here is not to perform the best component separation possible for the given experimental setup, but rather study the effect of the beams, we do not explore all possible options but, instead, choose one, which is reasonable given what is shown in~\cite{litebird2023probing}.
We thus consider the spatial variability values of \texttt{nside}~=~64 for $\beta_{\mathrm{dust}}$, \texttt{nside}~=~8 for $\mathrm{T}_{\mathrm{dust}}$, and \texttt{nside}~=~4 for $\beta_{\mathrm{synch}}$.

\begin{figure}
    \centering
    \includegraphics[width=\columnwidth]{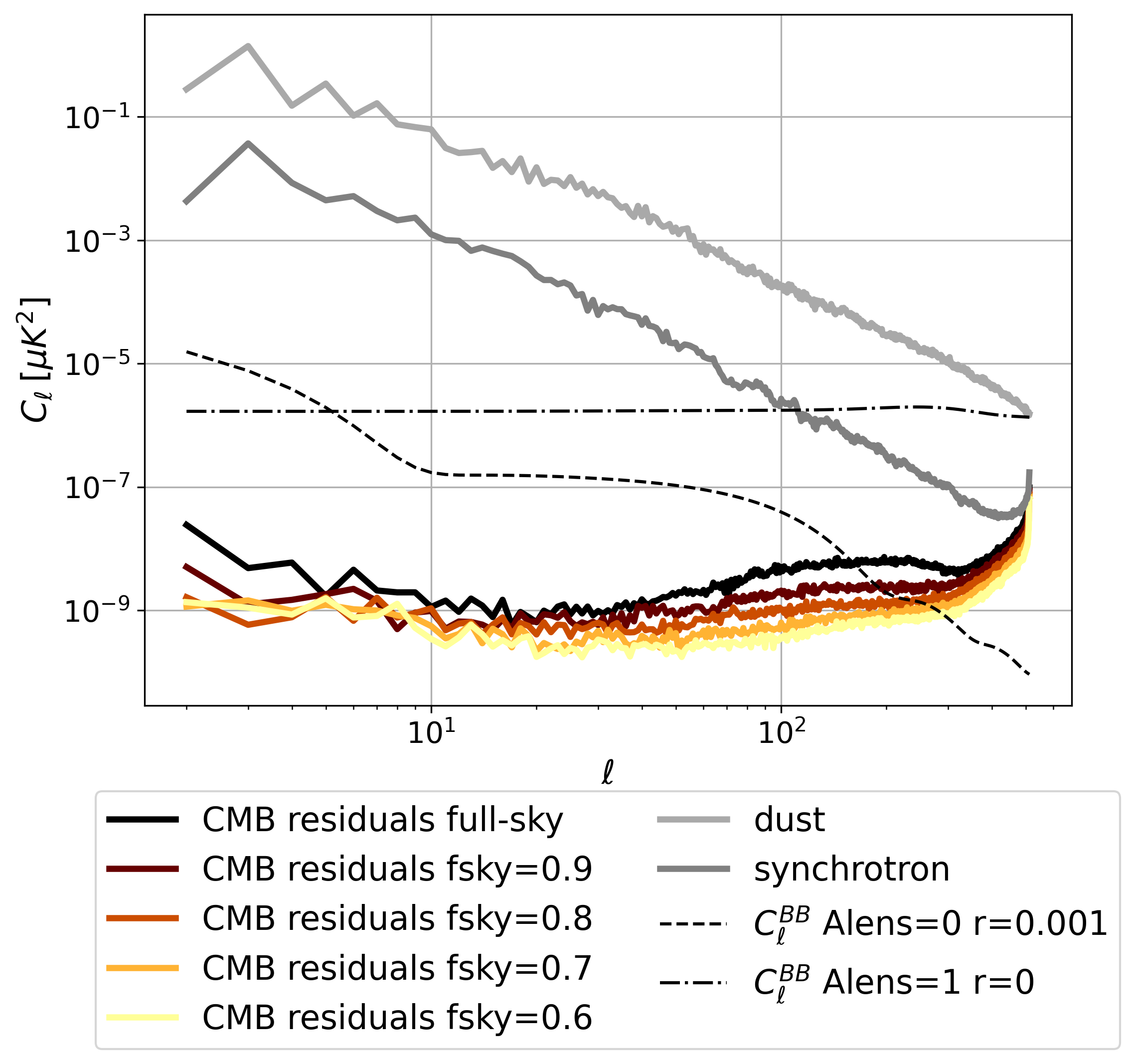}
    \caption{$BB$ CMB residuals due to the commutation of the mixing matrix and the beam operator $\mathbf{B}^{\phantom{-1}}_\mathbf{final}$ in the BIC-com approach, for different Galactic plane masks. The level of the residuals due to the beam commutation is here lower than how it is for the approach with the common resolution beam, thanks to the smaller beam that we can use in this approach.}
    \label{fig:w_commutation_AB_noiseless}
\end{figure}

We perform here only the second step of the component separation with the beams, assuming the spectral parameter values are given.
First we solve the PCG in the noiseless case, with the true spectral parameter values in one patch per pixel, to verify if we are able to recover the components as expected.
For the BIC-Ncom data model we can recover the correct components with numerically zero residuals, this is true for any FWHM value of $\mathbf{B}^{\phantom{-1}}_\mathbf{final}$, even if we neglect it altogether as in the data model in Eq.~(\ref{data_model_beams}), and we do it assuming a $\ell_\mathrm{max}$ value of about 300 as this is where the residuals will start to increase significantly due to the convolved noise.
In contrast, this is not the case for the data model BIC-com, see Figure~\ref{fig:w_commutation_AB_noiseless}, in which the residuals due to the commutation, for different fraction of the sky are shown.
The residuals that we get in this latter case give us the approximate measure of the error introduced assuming the commutation of $\mathbf{A}$ and $\mathbf{B}^{\phantom{-1}}_\mathbf{final}$ with this particular experimental setup.
We note however that the level of the residuals is much smaller than the targeted signal, which means that with this particular instrumental setup and amount of spatial variability performing the commutation is an acceptable approximation.

The value of $\mathbf{B}^{\phantom{-1}}_\mathbf{final}$ has been chosen to be $17'$, hence as the smallest of the true beams. In fact we find that the choice of the final beam size is largely irrelevant in the cases studied in this work, to the extent that we can neglect it altogether. Instead, a care has to be taken about the band-limiting of all the pixel domain objects. This may or may not be the case for more complex simulations and real data, where subpixel power aliasing may become important.
We therefore believe that using the explicit final beam is always prudent. 
We also find that it has an effect on the performances of the PCG.

\begin{figure}
    \centering
    \includegraphics[width=\columnwidth]{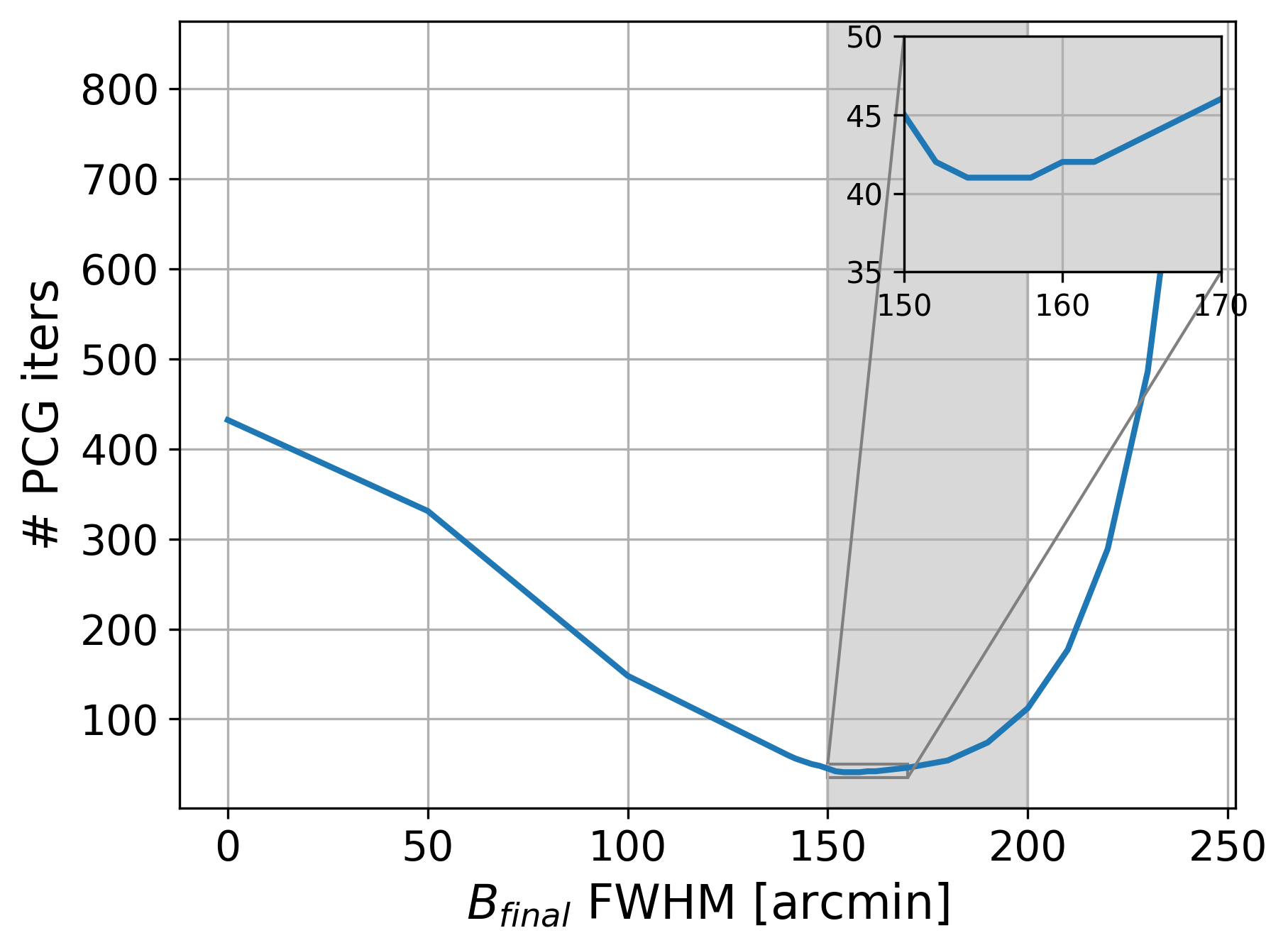}
    \caption{Number of PCG iterations needed to recover the components with different choices of $\mathbf{B}^{\phantom{-1}}_\mathbf{final}$. 
    Example for maps at \texttt{nside}~=~64 and with $\mathbf{B}^{\phantom{-1}}_\mathbf{true}$ FWHM in the range from $150'$ to $200'$. The minimum number of iterations required is for a $\mathbf{B}^{\phantom{-1}}_\mathbf{final}$ FWHM value inside the range of values for $\mathbf{B}^{\phantom{-1}}_\mathbf{true}$ and closer to its lower bound.}
    \label{fig:choice_Bfinal}
\end{figure}

We have studied how the number of required PCG steps varied to reach the tolerance of $10^{-7}$ as a function of the $\mathbf{B}^{\phantom{-1}}_\mathbf{final}$ chosen.
The qualitative behaviour that we have found does not change with the chosen \texttt{nside} or beam value: the minimum of iterations needed is for a beam width in between the values of the true beams, closer to the lower bound, see Figure~\ref{fig:choice_Bfinal}.
For smaller values the number of iterations needed significantly increases, the extreme case is the one of no $\mathbf{B}^{\phantom{-1}}_\mathbf{final}$, equivalent to the data model of Eq.~(\ref{data_model_beams}).
While on the other hand for larger beams the increase in number of iterations is even steeper.
We explain this behaviour as a trade-off between having badly constrained modes (too small beam size) or strongly correlated modes (too large beams) both negatively affecting the convergence.
Hence, having both $\mathbf{B}^{\phantom{-1}}_\mathbf{true}$ and $\mathbf{B}^{\phantom{-1}}_\mathbf{final}$ with similar values allows to have better numerical stability.
We show an example of this behaviour for a case at \texttt{nside}~=~64 and true beams ranging from $150'$ to $200'$.
We also notice that, given a particular $\mathbf{B}^{\phantom{-1}}_\mathbf{final}$, the data model BIC-com, converges faster than the one without commutation, BIC-Ncom, requiring about a factor 5 less iterations.

\begin{figure}
    \centering
    \includegraphics[width=\columnwidth]{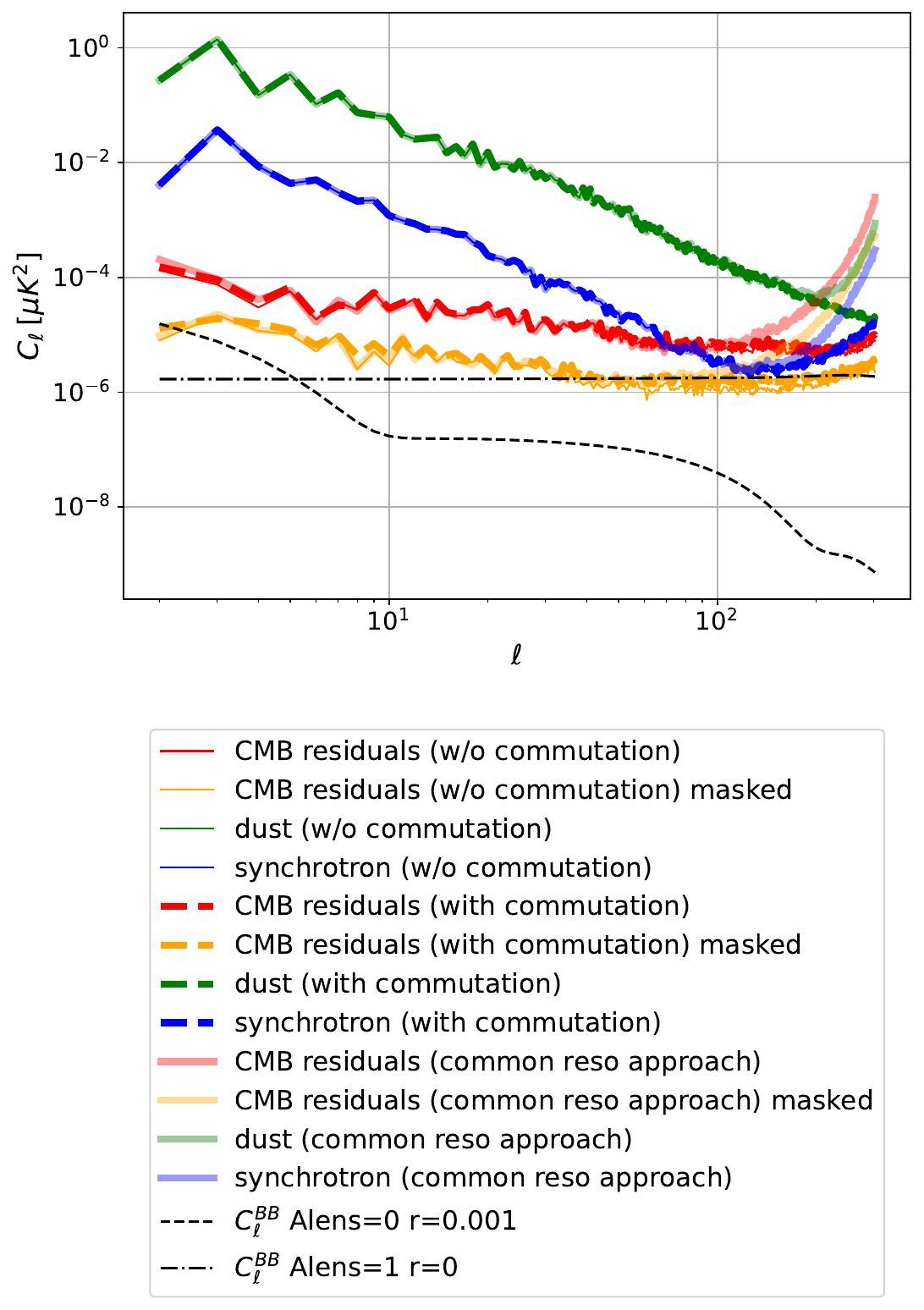}
    \caption{$BB$ spectra of the recovered components (100~GHz) for sky $d1s1$, $LiteBIRD$-like, full-sky. The CMB residuals are shown both full-sky (red lines) and partial-sky (yellow lines, Galactic plane masked with $f_{\mathrm{sky}}=0.4$). 
    The first step of the component separation is performed with the CRA, while the second step both with the CRA and with the beams in the component separation with both data model proposed, BIC-Ncom and BIC-com.
    In the latter two cases the recovered components are in very good agreement and are plotted here up to the scales at which the noise starts dominating. The smaller scales are instead out of reach when performing the second step with the CRA.
    The CMB residuals could be further lowered by optimizing both the patches to deal with the foreground scaling spatial variability and the Galactic plane mask.}
    \label{fig:recovered_components_w_betas_from_Beff_run}
\end{figure}

We now perform the second step of the component separation assuming the spectral parameter values as recovered from a run with the CRA.
As compared to the case without spatial variability, we now choose to downgrade to \texttt{nside}~=~64 instead of 32, to reduce the bias due to the downgrade in presence of spatially varying spectral parameters.
We compare the recovered components with the CRA and with the two data models, BIC-Ncom and BIC-com, see Figure~\ref{fig:recovered_components_w_betas_from_Beff_run}.
We can here make similar considerations to what is discussed in the case without spatial variability.
In particular with the data models with the beams we can reach smaller scales before the recovered residuals increase because dominated by the noise convolved to a combination of the $\mathbf{B_{true}}$ values, with respect to the CRA.
Again, a much wider range of $\ell$ is now accessible for the thermal dust component, while less for the synchrotron component, according to the frequency bands where the gain on the beams resolution is more important and the noise level of the instrument.
We also notice that the residual is now not fully dominated by the noise, as it was in the case without spatial variability, but are instead mainly dominated by systematic and statistical errors on spectral parameter estimates.
The noise starts dominating only at smaller angular scale, where the residuals go up.
Those scales, from about $\ell>100$, as are the ones affected by the beams, are the ones which require more PCG iterations to converge in the BIC approaches.
Indeed, on large scales, thanks to the preconditioner used, good convergence is reached after few iterations ($\mathcal{O}(1)$), while all the following iterations are needed to bring the final scales up as due to the beams.
The number of iterations needed strongly depends on the values used in the mixing matrix.
When using the true spectral parameter values used also to create the simulations in this experimental setup with \texttt{nside}~=~512, about 40 iterations are needed, while more and more iterations are required as the spectral parameters differ from these values.
In particular with the spectral parameter used here from the common resolution run in the multiresolution configuration described above, with relative small \texttt{nside} the number of iterations required is of order of tens, while it significantly increases with \texttt{nside}~=~512, up to order of thousands to reach a PCG tolerance of $10^{-6}$.
To reduce the number of iterations and make the approach less computationally expensive we decrease the $\ell_\mathrm{max}$ to which we recover the components.
This is because the smaller angular scales are the ones mostly affected by the beam and thus requiring more iterations to converge.
Typically it is enough to keep the scales up to where the noise starts dominating and exploding.
Limiting the $\ell$ range we consider also has the benefit of reducing the computational cost of the SHTs.

On a practical note, to allow for the PCG convergence one should ensure that the RHS and LHS of Eq.~(\ref{eq:compsep_step1}) have the same band limits. This is particularly the case with the generalized mixing matrix, Eq.~(\ref{eq:genMixingNonComm}) BIC-Ncom, as it requires the application of the beam operator as the last operation, which imposes a band limit when going back from harmonic to pixel domain, while it may not be the case for the data model of Eq.~(\ref{eq:genMixingComm}) BIC-com, for which then an additional, explicit bandlimiting of the RHS and LHS after the application of $\mathbf{A}$ is required.

\bigskip 

\subsection{Small sky fractions}
As a study case here we consider a \textit{Simons Observatory}-like setup, and in particular its small aperture telescopes, with the specifics given in Table~\ref{tab:SO-setup}~\cite{ade2019simons}.
\begin{table}[h]
\centering
\begin{tabular}{|c|c|c|c|}
\hline
\multicolumn{1}{|c|}{$\nu$} & sensitivity & FWHM \\
\multicolumn{1}{|c|}{[GHz]} & $[\mu \mathrm{Karcmin}]$ & [arcmin] \\
\hline
27.0  & 49.5  & 91.0  \\
39.0   & 29.7  & 63.0  \\
93.0    & 3.7   & 30.0  \\
145.0   & 4.7   & 17.0  \\
225.0   & 8.9   & 11.0  \\
280.0  & 22.6  & 9.0   \\
\hline
\end{tabular}
\caption{Frequency, sensitivity and FWHM of Gaussian beams for the $SO$-like setup~\cite{ade2019simons}.}
\label{tab:SO-setup}
\end{table}
We perform the second step of the component separation with the signal only coming from about 10\% of the sky.
Given the partial sky our method is now only approximately valid.
In the following we want to assess the impact of this approximation.
We notice that an error due to the cut-sky is also inevitably introduced in the CRA, as edge effects are present whenever we smooth the maps to the common resolution.

We perform the second step of the component separation procedure on $d1s1$ noiseless simulations (\texttt{nside}~=~512 and $\ell_\mathrm{max}$ for the SHTs in the PCG 1024), assuming their true spectral parameters.
This choice allows us to assign the full residuals that we get to the error introduced by the SHTs with the partial sky, as on full sky we would get numerically zero residuals.
For numerical efficiency we set the beamsize of the beams smaller than $17'$ to $17'$.

\begin{figure*}
    \centering
    \includegraphics[width=0.95\textwidth]{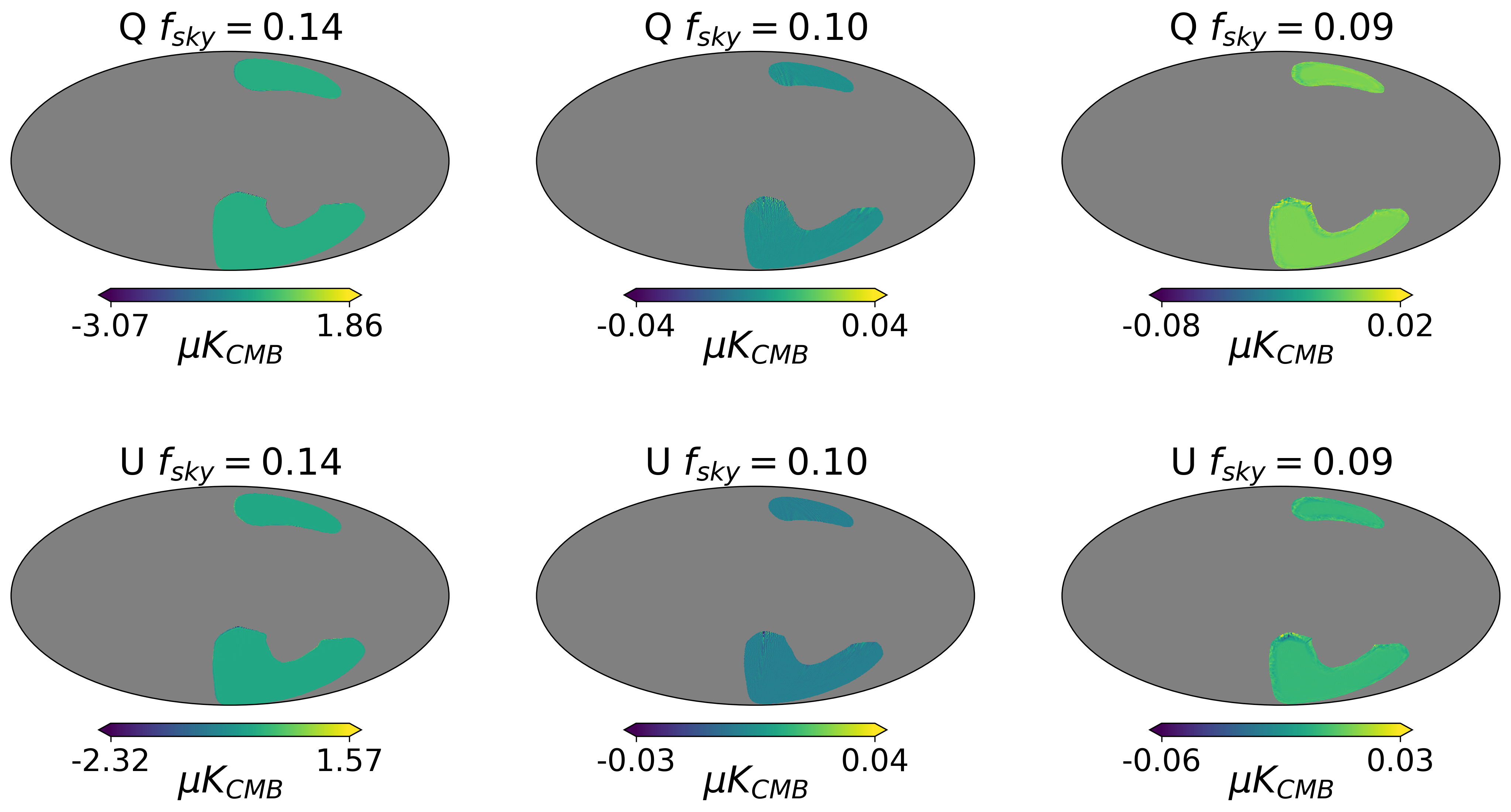}
    \caption{Recovered CMB residuals for an $SO$-like patch, $d1s1$ noiseless sky and with true spectral parameter values, the recovered CMB residuals are thus due to the error introduced by performing the SHTs on the cut-sky. In the first column the majority of the residual is in the pixels along the patch border, those have been masked in the central column, while on the right column the PCG has been solved on the apodized patch (with effective $f_{\mathrm{sky}}\approx 0.09$). Both these two latter cases show residuals by about an order of magnitude weaker than the former case.}
    \label{fig:maps_SO_1}
\end{figure*}

\begin{figure}
    \centering
    \includegraphics[width=\columnwidth]{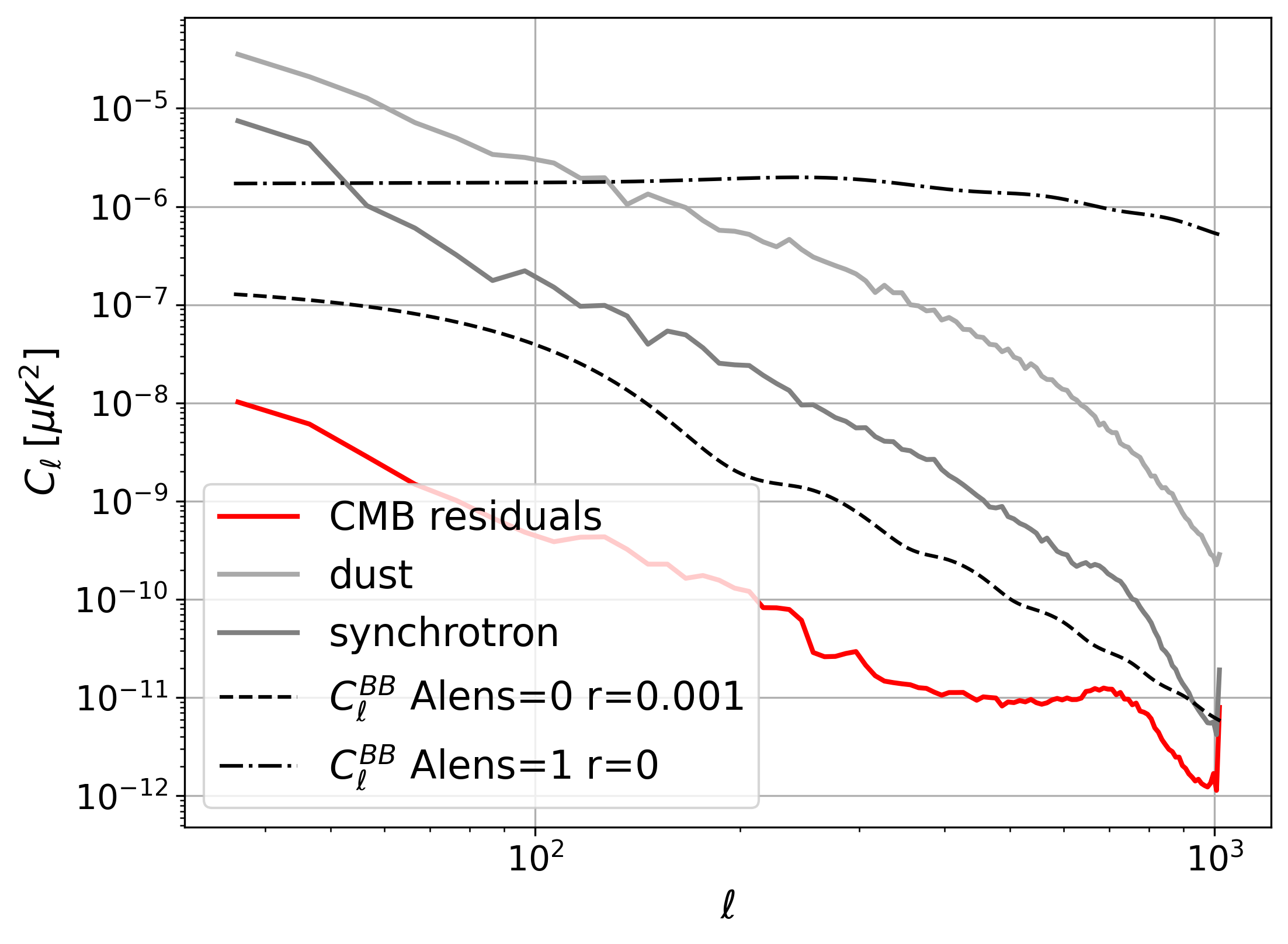}
    \caption{Cut-sky effect in the BIC method.
    We show the $BB$ power spectrum from the components reconstructed with the data model of Eq.~(\ref{data_model_beam1}) and the PCG with tolerance $10^{-6}$, with the apodized mask with effective $f_\mathrm{sky} \approx 0.09$. The components are plotted for the frequency of 100~GHz. The input sky is $d1s1$ noiseless and the spectral parameters used are the true ones, the recovered CMB residuals are thus due to the error introduced by performing the SHTs on the cut-sky. They are lower than the targeted signal and they could be further reduced by removing the pixels on the border of the patch.}
    \label{fig:Cls_SO}
\end{figure}

The recovered components are shown in Figure~\ref{fig:maps_SO_1}.
The main bias is visible along the patch border due to the signal being zero inside the mask.
We can easily remove those biased pixels at the cost of reducing the sky fraction.
We also notice that for an apodized initial mask the effect of the border is faded.
We show the recovered $B$-mode residual power spectrum for this latter case, Figure~\ref{fig:Cls_SO}. 
Given that the error due to the mask is limited and not impacting the scientific target the proposed method can be used in this study case too, with analogous gain in resolution to what was described in the previous section.
At the cost of losing numerical efficiency one could keep the lower resolution channels to their native values and performing the computation at \texttt{nside}~=~1024, which would lead to even further gain in resolution.

\section{Conclusion and prospects}\label{sec:conclusion}
In this work we have discussed a treatment of main beam in the context of the parametric component separation and the input single frequency maps characterized by different observational beams. We have considered the standard approach, referred to as a CRA, where all the input maps are smoothed to the common resolution prior to the component separation. We have found out that this approach does not only tend to produce results with lower resolution but also that it implicitly assumes commutations of the common beam operator and mixing matrix. 
We have shown that this is the case even for mildly spatially-varying foreground properties, and this unavoidably leads to a bias, which can, and has to, be controlled with appropriate masking of the high contrast regions, where the variability of the foreground properties is particularly important. 
While we have found the bias to be manageable in the examples studied here, it depends strongly on the scale of the smoothing and thus it increases for large beams, becoming progressively more important and/or requiring more aggressive masking. We have then proposed two alternative approaches where the beams are explicitly treated as part of the component separation procedure. They either do not assume the beam and mixing matrix commutation at all, BIC-Ncom,
or assume but only for beams much finer that the ones typically required for the CRA, BIC-com.
We have developed numerically viable implementations of both these methods, which are algorithmically much more involved than the CRA. We have found out that indeed these new methods allow us to recover sky signals over a significantly broader range of angular scales than the CRA and can do so while controlling much better the level of beam-induced residuals. We have demonstrated both these methods on nearly full-sky simulations as expected from the next generation satellite missions and on the limited sky observations as could be observed from the ground. We have found out that the boundary effects due to the cut sky can be successfully dealt with. From the two proposed methods we have found that the method which does not assume any beam-mixing matrix commutation, BIC-Ncom, tends indeed to produce the most accurate results, however, this comes at the cost of significant numerical load, as the PCG solver in this case converges very slowly. Instead, the simpler method, which assumes the commutation of a high resolution beam and the mixing matrix, BIC-com, seems to offer an interesting trade-off between the computational efficiency and the level of bias, ensuring that the latter is subdominant for the sensitivity levels expected for the next generation of experiments and that the solution is produced in reasonable computation time. In both these cases further algorithmic improvements are however still needed to allow for a full exploration of the relevant likelihoods. This is a target of the on-going work.

The obtained results are very promising and provide a proof-of-concept for this type of approaches, which will be instrumental in facilitating the full exploitation of the scientific potential of the future CMB data sets.
The current implementation is however limited to axially symmetric beams and either spatially correlated stationary noise or inhomogeneous, uncorrelated noise. All these limitations can, and should, be bypassed with new, more efficient numerical algorithms and their implementations. Furthermore, a full treatment of the problem will also require an inclusion of instrumental systematic effects. We leave such extensions for future work noting only here that all these effects are also relevant for the CRA, where their impact also has to be yet assessed and fully understood.

\section*{Acknowledgments}
The authors thank Dominic Beck and Hamza El Bouhargani for useful discussions in the initial phases of the project.
AR also thanks Magdy Morshed for various feedback during the project.
Some of the results in this paper have been derived using the \texttt{healpy}~\cite{Zonca2019} and \texttt{HEALPix} package~\cite{2005ApJ...622..759G} and with the \texttt{NaMaster} package \cite{alonso2019unified}.
The authors acknowledge the SCIPOL project funded by the European Research Council (ERC) under the European Union’s Horizon 2020 research and innovation program (Grant agreement No. 101044073). This work has also received funding by the European Union’s Horizon 2020 research and innovation program under grant agreement no. 101007633 CMB-Inflate.
This work was granted access to the HPC resources of IDRIS under the allocation 2024-AD010414161R1 made by GENCI.
Numerical computations were partly performed on the DANTE platform, APC, France.

\bibliographystyle{abbrv}
\bibliography{bibliography.bib}

\end{document}